\documentclass[aps,prd,preprint,groupedaddress]{revtex4-2}
\usepackage{graphicx}
\usepackage{amsmath}
\usepackage{amssymb}
\usepackage{hyperref}
\usepackage{bm}

\let\l=\left
\let\r=\right
\def\be{\begin{equation}}
\def\ee{\end{equation}}
\def\bea{\begin{eqnarray}}
\def\eea{\end{eqnarray}}

\DeclareMathOperator\arctanh{Arctanh}

\begin{document}

%\preprint{}

\title{Boundary conditions for stationary black holes ; Application to Kerr, Martinez-Troncoso-Zanelli and hairy black holes.}


\author{Philippe Grandcl\'ement}
\email[]{Philippe.Grandclement@observatoiredeparis.psl.eu}
\author{Jordan Nicoules}
\email[]{Jordan.Nicoules@observatoiredeparis.psl.eu}
%\thanks{}
\affiliation{Laboratoire Univers et Th\'eories, Observatoire de Paris, Universit\'e PSL, Universit\'e Paris Cit\'e, CNRS, F-92190 Meudon, France}


\date{\today}

\begin{abstract}

This work proposes a set of equations that can be used to numerically compute spacetimes containing a stationary black hole.
The formalism is based on the 3+1 decomposition of General Relativity with maximal slicing and spatial harmonic gauge. 
The presence of the black hole is enforced using the notion of apparent horizon in equilibrium. 
This setting leads to the main result of this paper: a set of boundary conditions describing the horizon and that must be used when solving the 3+1 equations. 
Those conditions lead to a choice of coordinates that is regular even on the horizon itself. 
The whole procedure is validated with three different examples chosen to illustrate the great versatility of the method. 
First, the single rotating black holes are recovered up to very high values of the Kerr parameter. 
Second, non-rotating black holes coupled to a real scalar field, in the presence of a negative cosmological constant (the so-called MTZ black holes), are obtained. 
Last, black holes with complex scalar hairs are computed. 
Eventually, prospects for future work, in particular in contexts where stationarity is only approximate, are discussed.

\end{abstract}

%\keywords{}

\maketitle

\section{\label{s:intro} Introduction}

Black holes are objects so compact that nothing, light included, can escape their intense gravitational field. 
If this concept was first considered in the eighteenth century, it is with the advent of General Relativity that the mathematical description of those objects was made possible. 
The first solution of a single non-rotating black hole was obtained by Schwarzschild in 1916.
In 1963, Kerr extended the solution to include rotation and the metric which took his name \cite{Kerr:1963}. 
Not only does this metric describe a rotating black hole, but it has been proven that, under some assumptions, it is the only possible choice. 
This result arises from the uniqueness theorems (see \cite{Chrusciel:2012} for a review).

If it is long known that massive stars should end up their life producing a black hole, direct proofs of the existence of astrophysical black holes are now available with more and more confidence. 
First there is the detection of the gravitational waves emitted by the coalescence of two black holes. 
When two such objects are orbiting each other, they lose energy by deforming the spacetime and eventually merge into a single black hole. 
The emitted gravitational waves can be detected by laser interferometry on Earth \cite{LIGO, Virgo}. 
Since the first detection in 2015 \cite{Abbott:2016}, several tens of such binaries have been detected. 
The observed waveforms are in total agreement with the prediction of General Relativity \cite{LIGOScientific:2019, LIGOScientific:2021}.

On the other hand, observations with very high angular resolutions enabled to confirm the existence of supermassive black holes at the center of galaxies, mainly in two cases. 
First, at the center of our galaxy where the orbits of stars indicate the existence of a small dark object of mass $\approx 4 \cdot 10^{6} M_\odot$. 
Observations are consistent with this object being a classical black hole described by General Relativity \cite{GRAVITY:2020}. 
Using very-long baseline interferometry it was also possible to directly image an accretion disk around the center of the galaxy M87 \cite{EHT:2019}. 
The obtained image exhibits a shadow consistent with the fact that the disk is indeed orbiting a black hole.

It is expected that all those types of detectors will reach higher and higher sensitivity in the coming years. 
New detectors, like the LISA space interferometer \cite{LISA, Amaro-Seoane:2022} or the Einstein Telescope \cite{Maggiore:2019}, will also come online .
This will enable precise tests of the black hole paradigm. 
If so far observations are consistent with the compact objects being classical Kerr black holes, there is possibility that they are indeed more complicated ones. 
An example of such an alternative model is the black hole with scalar hairs studied in Sec. \ref{s:hairy}. 
To maximize the scientific impact of the future observations, theoretical studies of the various models of black holes are needed. 
A lot of those studies rely on analytic choices, in particular concerning the coordinates used. 
This can be a limitation when those coordinates are singular or when fast rotation is included (see \cite{VanAelst:2019} for instance).

In this paper, a framework is proposed that enables the numerical description of stationary black holes in a rather general context. 
The formalism should lead to a choice of coordinates that is regular everywhere, in particular across the black hole horizon. 
The formalism relies on the 3+1 decomposition of spacetime. 
The choice of coordinates is based on maximal slicing for the time coordinate and on the spatial harmonic gauge for the spatial ones. 
The presence of the black hole is enforced by demanding that spacetime contains an apparent horizon in equilibrium. Let us mention that this is far from
being the first proposal for such a description. A method based on the spatial symmetries of the spacetimes can be found in \cite{Headrick:2010, Adam:2012, Wiseman:2012}.
It has been applied successfully to various situations (asymptotically anti-de Sitter, various dimensions). A method relying on the 
use of the Dirac gauge can be found in \cite{Vasset:2009} where it is applied to the Kerr spacetime.

The paper is organized as follows. 
In Sec. \ref{s:equations}, the formalism is presented. 
The bulk equations coming from the 3+1 setting are exhibited. 
The description of the apparent horizon is also investigated in detail and the resulting boundary conditions for the metric fields are given. 
Three different applications are then shown. 
In Sec. \ref{s:kerr}, the Kerr black hole is recovered in this coordinate system (which is not analytic). 
In Sec. \ref{s:MTZ}, a model of black hole studied by Martínez and collaborators is recovered numerically. 
The model has no angular dependence and contains a real scalar field minimally coupled to gravity. 
There is also a negative cosmological constant which causes the spacetime to be asymptotically anti-de Sitter. 
Section \ref{s:hairy} shows the construction of black holes with complex scalar hairs as already obtained in \cite{Herdeiro:2014, Herdeiro:2015}. 
Future prospects are considered in Sec. \ref{s:conclusions}.

Throughout this paper Greek indices are four-dimensional ones, ranging from 0 to 3 whereas Latin indices are spatial ones, ranging from 1 to 3. Units such that $G=c=1$ are used.

\section{\label{s:equations} Formalism}
\subsection{\label{ss:gauge} 3+1 formalism and gauge conditions}

The 3+1 decomposition of Einstein's equations is widely used in the field of numerical relativity and it is at the core of this work as well. 
Basic features of this formalism are recalled (see for instance \cite{Gourgoulhon:2007} for more details). 
The four-dimensional metric ${\bf g}$ is decomposed as
\be
g_{\mu\nu} {\rm d}x^\mu {\rm d}x^\nu = (-N^2 + B_i B^i) {\rm d}t^2 + 2 B_i {\rm d}x^i {\rm d}t + \gamma_{ij} {\rm d}x^i {\rm d}x^j.
\ee

The hypersurfaces $\Sigma_t$ of constant time $t$ are mapped by the purely spatial coordinates $x^i$. 
The 3+1 quantities are then a scalar function $N$ the lapse, a vector field $B^i$ the shift and the metric induced on $\Sigma_t$, $\gamma_{ij}$. 
All indices of spatial quantities are manipulated by the induced metric. 
In the following $\nabla$ denotes the covariant derivative associated with $g_{\mu\nu}$ and $D$ the one associated with $\gamma_{ij}$.

The normal to each slice $\Sigma_t$ is $n_\mu = \l(-N, 0, 0, 0\r)$. In this framework, the second fundamental form, the extrinsic curvature tensor, reads as follows:
\be
\label{e:extrinsic}
K_{ij} = \frac{1}{2N} \l(D_i B_j + D_j B_i - \partial_t \gamma_{ij}\r).
\ee

Each index of Einstein's equations can then be projected either on the hypersurface $\Sigma_t$ or along the normal. 
It leads to the the 3+1 equations of general relativity:

\bea 
\label{e:hamilton}
H &:& R + K^2 - K_{ij} K^{ij} - 2 \Lambda = 16 \pi E \\
\label{e:momentum}
M_i &: & D^j K_{ij} - D_i K = 8 \pi P_i \\
\label{e:evol}
E_{ij} &:& -\partial_t K_{ij} + \mathcal{L}_{\bm{B}} K_{ij} - D_i D_j N + N \l(R_{ij} + K K_{ij} - 2 K_{ik} K_j^k - \Lambda \gamma_{ij}\r) \\
\nonumber &=& 4 \pi  N  \l( 2 S_{ij} - \l(\gamma^{kl} S_{kl} - E\r) \gamma_{ij}\r) ,
\eea
where $\mathcal{L}$ denotes the Lie derivative, $R_{ij}$ and $R$ the Ricci tensor and scalar and $K$ the trace of the extrinsic curvature tensor.
$H$ denotes the Hamiltonian constraint, $M_i$ the momentum constraint and $E_{ij}$ the evolution equation. The equations are written here with a cosmological constant $\Lambda$ and in the presence of matter. Matter terms contain the 3+1 projections of the stress-energy tensor $E$, $P_i$ and $S_{ij}$.

As such, the system of equations (\ref{e:hamilton}-\ref{e:evol}) (supplemented with Eq. (\ref{e:extrinsic})), cannot be solved to find the fields $N$, $B^i$ and $\gamma_{ij}$. 
Indeed the choice of coordinates $\l(t, x^i\r)$ has not been prescribed yet and the general covariance of the theory would lead to an ill-posed problem.

In the following, one uses the same choice of coordinates that was successfully employed in \cite{Martinon:2017} and which mathematical properties were assessed in \cite{Andersson:2001}. 
The slicing of spacetime (i.e. the choice of $\Sigma_t$) is defined by the maximal slicing condition $K=0$. 
One also demands that the spatial coordinates fulfill the spatial harmonic gauge (the 3D version of the well-known 4D harmonic one). 
It amounts to enforcing that 
\be
\label{e:harmonic}
V^k \equiv \gamma^{ij} \l(\Gamma^k_{ij} - \bar{\Gamma}^k_{ij}\r) = 0.
\ee
$\Gamma^k_{ij}$ are the Christoffel's symbols of $\gamma_{ij}$ whereas the $\bar{\Gamma}^k_{ij}$ correspond to a fixed background metric 
$\bar{\gamma}_{ij}$. 
Different choices of the background metric would lead to different choices of spatial coordinates. 
In standard cases the background metric is usually the flat metric one (in spherical or Cartesian coordinates). 
When a negative cosmological constant is present (as in \cite{Martinon:2017} or in Sec.~\ref{s:MTZ}) it is convenient to consider spatial metrics linked to the anti-de Sitter (ADS) spacetime.

In \cite{Andersson:2001} it has been shown that the Ricci tensor can be expressed as 
\be
R_{ij} = -\frac{1}{2} \gamma^{kl} \bar{D}_k \bar{D}_l \gamma_{ij} + \frac{1}{2} \l(D_i V_j + D_j V_i\r) + {\rm terms} (\partial\gamma ~\partial\gamma),
\ee
where $\bar{D}$ denotes the covariant derivative associated with the background metric. Using $V^i=0$ thus ensures that the second order derivatives of the metric appearing in the Ricci tensor are all accounted for by $\gamma^{kl} \bar{D}_k \bar{D}_l \gamma_{ij}$, which is a Laplacian-like operator and thus well-behaved (see Sec. (3.2) of \cite{Martinon:2017} for more details).

In order to enforce the gauge conditions, one removes all the occurrence of $K$ in the 3+1 equations.
$R_{ij}$ is also replaced by $R_{ij} - 1/2 \l(D_i V_j + D_j V_i\r)$, making the associated second order part Laplacian-like. 
In the following, only stationary black holes are considered so that one can also remove all the terms $\partial_t$. 
This leads to the following system of equations:

\bea
\label{e:hamilton_ok}
H &:& R - D_k V^k - K_{ij} K^{ij} - 2 \Lambda = 16 \pi E \\
\label{e:momentum_ok}
M_i &:& D^j K_{ij} = 8 \pi P_i \\
\label{e:evol_ok}
E_{ij} &:& \mathcal{L}_{\bm{B}} K_{ij} - D_i D_j N + N \l(R_{ij} - \frac{1}{2} \l(D_i V_j + D_j V_i\r) - 2 K_{ik} K_j^k - \Lambda \gamma_{ij}\r) \\
\nonumber
&=& 4 \pi  N  \l( 2 S_{ij} - \l(\gamma^{kl} S_{kl} - E\r) \gamma_{ij}\r) ,
\eea
where $V^i$ is given by Eq. (\ref{e:harmonic}) and $K_{ij}$ by Eq. (\ref{e:extrinsic}), which in that case reduces to:
\be
\label{e:extrinsic_ok}
K_{ij} = \frac{1}{2N} \l(D_i B_j + D_j B_i\r).
\ee

Eqs. (\ref{e:hamilton_ok}-\ref{e:evol_ok}) are now an invertible system of 10 components, corresponding to the 10 unknown fields $N$, $B^i$ and $\gamma_{ij}$.
Once this system is solved, a very important check consists in verifying, {\em a posteriori}, that the gauge fields $K$ and $V^i$ are indeed zero. 
If this is not the case, the solved equations do not coincide with the original ones, making the overall procedure fail. 
This check is of uttermost importance and it is very difficult to pass if anything is wrong somewhere in the equations. 
It is carefully monitored for the three different examples presented in this paper. The above procedure is the three-dimensional equivalent 
of the so-called De Turck procedure used to enforce the four-dimensional harmonic gauge \cite{Deturck:1983, Headrick:2010}. Validity of the procedure
is discussed, for instance, in \cite{Figueras:2011, Figueras:2017}.

\subsection{Apparent horizon boundary conditions}

It is well-known that event horizons are global objects that require the knowledge of the full spacetime in order to be located. 
It makes them difficult to use in the context of numerical relativity. 
In order to enforce the presence of black holes, one usually relies on the local notion of apparent horizon, first introduced in  \cite{Hawking:1973}. 
By local one means that apparent horizon can be defined on each slice $\Sigma_t$ by the sole knowledge of the geometry of the slice. 
Apparent horizons are commonly used in black hole simulations (see for instance \cite{Gourgoulhon:2001, Grandclement:2001, Caudill:2006} for applications in the binary context). 
In the following only relevant properties of apparent horizons are discussed (see \cite{Gourgoulhon:2005} for a detailed review). 
Let us also point out that, in the stationary cases, apparent and event horizon coincide.

Without loss of generality, the apparent horizon is assumed to be a sphere of constant radius $r_{\rm H}$. 
Spherical coordinates $\l(r,\theta,\varphi\r)$ are used to map the slices $\Sigma_t$ and tensors are given with respect to the associated (orthonormal) spherical tensorial basis. 
Let us denote $\tilde{s}^i$ the unit, outward, spacelike, normal to the horizon. If need be, it can be shown that
\bea
\tilde{s}_i &=& \l(\sqrt{\gamma^{rr}}, 0, 0\r) \\
\tilde{s}^i &=& \frac{1}{\sqrt{\gamma^{rr}}} \l(\gamma^{rr}, \gamma^{r\theta}, \gamma^{r\varphi}\r).
\eea
The very definition of an apparent horizon means that it is the outermost trapped surface. 
So the expansion $\Theta$ of the outward future null vector must vanish. 
Using the 3+1 expression for $\Theta$ (see \cite{Gourgoulhon:2005}) it gives rise to the following equation
\be
\label{e:expansion}
\Theta \equiv D_i \tilde{s}^i + \tilde{s}^i  \tilde{s}^j K_{ij} = 0.
\ee

Moreover, following \cite{Caudill:2006, Gourgoulhon:2005}, one can ask that the coordinate system is stationary with respect to the horizon (i.e. the horizon location is fixed). 
It implies that, on the horizon
\be
\label{e:lapsehor}
N = B^i \tilde{s}_i.
\ee
One can also show (see for instance \cite{Vasset:2009}) that the shear $\sigma_{ab}$ of the null light rays must vanish. 
It implies that the shift vector $B^i$ must be a conformal Killing vector of the sphere. 
A possible choice together with Eq. (\ref{e:lapsehor}) is 
\be
\label{e:bc_shift}
B^i = N \tilde{s}^i - \Omega_{\rm BH} m^i,
\ee
where $\Omega_{\rm BH}$ is a constant that captures the rotation velocity of the black hole and $m^i = \l(\partial_\varphi\r)^i$. 
The minus sign ensures that the angular momentum is positive.

Equations (\ref{e:expansion}-\ref{e:bc_shift}) have been successfully used in many publications, especially for computing binary black holes initial data \cite{Caudill:2006, Uryu:2012, Papenfort:2021}. 
However in those papers the spatial metric is set to be conformally flat, an assumption that is not used in this work.

\subsection{Differential gauges}

Maximal slicing and the spatial harmonic gauges are differential gauges. 
By that, it is to be understood that they do not impose conditions directly on the metric fields but rather lead to partial differential equations.
When regularity of the full spacetime is required, it leads to a unique choice of coordinates. 
However, as will be illustrated below, when a horizon is present, the coordinate system is fixed up to some boundary conditions which can be used to freely specify some quantities.

Consider an infinitesimal coordinate change of time of the form $t' = t + \alpha\l(x^i\r)$. 
At first order it does induce a change on the lapse and the shift proportional to $\partial_i \alpha$. 
Demanding that the new coordinate system obeys the maximal slicing condition $K'=0$ leads to a partial differential equation which is second order in terms of the coordinate change $\alpha$, $K'$ containing first order derivatives of the shift. 
So in order to transform an arbitrary coordinate system into one with maximal slicing, one needs to solve a second order partial differential equation for $\alpha$. 
It follows that $\alpha$ is determined up to two boundary conditions. 
The one at infinity is implicitly accounted for by demanding that the metric takes a fixed form (flat one in Sec \ref{s:kerr} and \ref{s:hairy} or ADS in Sec. \ref{s:MTZ}). 
In the presence of a horizon, the inner boundary condition for $\alpha$ can translate in the free choice of the lapse. So on the horizon the lapse is a freely specifiable angular function $N_0\l(\theta,\varphi\r)$.

The situation concerning the spatial coordinates is similar. 
Considering a coordinate change of the form $x'^i = x^i + \xi^i\l(x^j\r)$, one can show that the spatial harmonic gauge equation (\ref{e:harmonic}) is of second order in terms of $\xi^i$. 
Indeed the Christoffel's symbols contain first order derivatives of the metric which, in turn, contains first order derivatives of $\xi^i$. 
However not all of the components of $\xi^i$ can be chosen freely. 
Remember that the location of the horizon has been chosen beforehand as being a sphere of constant radius. 
In order to maintain this location, one needs to have $\xi^r =0$. 
The two angular components of the coordinate change, however, can be freely chosen on the horizon and this choice translates into the possible free choice of some of the 3+1 quantities. 
The shift being fixed by Eq. (\ref{e:bc_shift}), it is more convenient to enforce the value of some components of the metric. 
The most natural choice is to fix the components $\gamma_{r\theta}$ and $\gamma_{r\varphi}$ of the spatial metric (with the spectral methods used in this paper those components have the same spectral bases as $\xi^\theta$ and $\xi^\varphi$, leading to a well posed numerical system ; this is probably more profound than just technicalities of the numerical method). 
So, on the horizon one sets $\gamma_{r\theta} = f$ and $\gamma_{r\varphi} = g$, where $f$ and $g$ are arbitrary angular functions.

\subsection{Degeneracy of the equations}\label{ss:dege}

In the context of this work, an equation is said to be degenerate if the prefactor of the highest order derivative vanishes. 
As an illustration, consider the following equation $a\l(x\r) f'' + b\l(x\r) f' + c\l(x\r) f + d\l(x\r) = 0$ on $\l[-1, 1\r]$. 
If $a\l(x\r)$ does not vanish, then it can be solved with the imposition of two boundary conditions, at $x=-1$ and $x=1$. 
If $a\l(x=-1\r)=0$, the equation is degenerate and one can no longer choose any boundary condition at $x=-1$. 
Indeed, at this point, the equation reduces to $ b\l(x\r) f' + c\l(x\r) f + d\l(x\r)=0$ which is the only compatible choice and must be used as a boundary condition. 
In a sense, the equation is its own boundary condition.

This type of behavior is present in the 3+1 equations considered here and it must be dealt with carefully. 
This can be seen in the evolution equations (\ref{e:evol_ok}) where the factor in front of the second order radial derivatives of the metric is $\l(\l(B^r\r)^2 - N^2\gamma^{rr}\r)/2N$ which vanishes on the horizon, given Eq. (\ref{e:lapsehor}). 
However, as the full set of equations is coupled, one can not simply assign one equation to one unknown. 
The system should be considered as a whole. 
In order to do so, one needs to isolate, in Eqs.~(\ref{e:hamilton_ok}-\ref{e:evol_ok}) the terms involving the second radial derivatives of the fields.

\be
\label{e:matrix}
\l(
\begin{array}{c|ccc|c}
0 & 0 & 0 & 0 & -\frac{1}{2} \gamma^{rr} \\
\hline
 0 & \frac{1}{2N} \l(1+\gamma^{rr}\gamma_{rr}\r) &  \frac{1}{2N} \gamma^{rr}\gamma_{r\theta} &  \frac{1}{2N} \gamma^{rr}\gamma_{r\varphi} &
\frac{1}{2N} \gamma^{rr} B^r \\
0 & \frac{1}{2N} \gamma^{r\theta}\gamma_{rr} &  \frac{1}{2N} \gamma^{rr}\gamma_{\theta\theta} &  \frac{1}{2N} \gamma^{rr}\gamma_{\theta\varphi} &
0 \\
0 & \frac{1}{2N} \gamma^{r\varphi}\gamma_{rr} &  \frac{1}{2N} \gamma^{rr}\gamma_{\theta\varphi} &  \frac{1}{2N} \gamma^{rr}\gamma_{\varphi\varphi} &
0 \\
\hline
-1 & \frac{1}{N} B^r \gamma_{rr} & \frac{1}{N} B^r \gamma_{r\theta} & \frac{1}{N} B^r \gamma_{r\varphi} & \frac{1}{2N} \l(\l(B^r\r)^2 - N^2\gamma^{rr}\r) \\
0 & \frac{1}{2N} B^r \gamma_{r\theta} & \frac{1}{2N} B^r \gamma_{\theta\theta} & \frac{1}{2N} B^r \gamma_{\theta\varphi} & 0 \\
0 & \frac{1}{2N} B^r \gamma_{r\varphi} & \frac{1}{2N} B^r \gamma_{\theta\varphi} & \frac{1}{2N} B^r \gamma_{\varphi\varphi} & 0 \\
 0 & 0 & 0 & 0 & 0 \\
 0 & 0 & 0 & 0 & 0 \\
 0 & 0 & 0 & 0 & 0 \\
\end{array}\r.
\ee

\be
\nonumber
\hspace{-0.5cm}
\l.
\begin{array}{ccccc}
 -\frac{1}{2} \gamma^{r\theta} & -\frac{1}{2} \gamma^{r\varphi} & -\frac{1}{2} \gamma^{\theta\theta} &
-\frac{1}{2} \gamma^{\theta\varphi} & -\frac{1}{2} \gamma^{\varphi\varphi} \\
\hline
\frac{1}{2N} \gamma^{r\theta} B^r & \frac{1}{2N} \gamma^{r\varphi} B^r & 0 & 0 & 0 \\
\frac{1}{2N} \gamma^{rr} B^r &  0 & \frac{1}{2N} \gamma^{r\theta} B^r & \frac{1}{2N} \gamma^{r\varphi} B^r & 0 \\
0 & \frac{1}{2N} \gamma^{rr} B^r &  0 & \frac{1}{2N} \gamma^{r\theta} B^r & \frac{1}{2N} \gamma^{r\varphi} B^r \\
\hline
0 & 0 & 0 &0 &0 \\
 \frac{1}{2N} \l(\l(B^r\r)^2 - N^2\gamma^{rr}\r) & 0 & 0 & 0 & 0 \\
  0 &
 \frac{1}{2N} \l(\l(B^r\r)^2 - N^2\gamma^{rr}\r) & 0 & 0 & 0 \\
 0 & 0 &  \frac{1}{2N} \l(\l(B^r\r)^2 - N^2\gamma^{rr}\r) & 0 & 0 \\
 0 & 0 & 0 & \frac{1}{2N} \l(\l(B^r\r)^2 - N^2\gamma^{rr}\r) & 0 \\
0 & 0 & 0 & 0 & \frac{1}{2N} \l(\l(B^r\r)^2 - N^2\gamma^{rr}\r) \\
\end{array}\r)
\ee

This can be done and the result is given by (\ref{e:matrix}). 
The lines correspond to the equations of the system (i.e. line 1 to Eq. (\ref{e:hamilton_ok}), line 2 to 4 to Eq. (\ref{e:momentum_ok}) and lines 5 to 10 to Eq. (\ref{e:evol_ok})). 
The columns correspond to the metric fields (i.e. column 1 to $N$, columns 2 to 4 to $B^i$ and columns 5 to 10 to $\gamma_{ij}$). 
The quantities in (\ref{e:matrix}) are then the factors of the terms $\partial^2_{rr}$ of a given field, in a given equation.

As already noted, the factor appearing on the last lines vanishes on the horizon. 
Though it is difficult to find explicitly the eigenvalues of (\ref{e:matrix}), one can investigate them numerically after assigning some random but realistic values to the fields on the horizon. 
It appears that the multiplicity of the null eigenvalue is always three. 
It follows that the only degenerate equations correspond to the last three lines of (\ref{e:matrix}), which are the purely angular components of Eq. (\ref{e:evol_ok}) (i.e. the components $\l(\theta,\theta\r)$, $\l(\theta,\varphi\r)$ and $\l(\varphi,\varphi\r)$). 
Those components must be solved without any boundary conditions or equivalently as being their own boundary conditions.

\subsection{Behavior of the expansion}\label{ss:expansion}

At this point, it seems that there are enough boundary conditions to solve the problem. 
Indeed, one could use $N=N_0$, $B^i = N \tilde{s}^i - \Omega_{\rm BH} m^i$, $\Theta=0$, $\gamma_{r\theta}=f$, $\gamma_{r\varphi}=g$ and $E_{\theta\theta}=E_{\theta\varphi} = E_{\varphi\varphi}=0$. 
This is a set of ten boundary equations for the ten unknown fields $N$, $B^i$ and $\gamma_{ij}$.

\begin{figure}
\includegraphics[width=.7\textwidth,keepaspectratio]{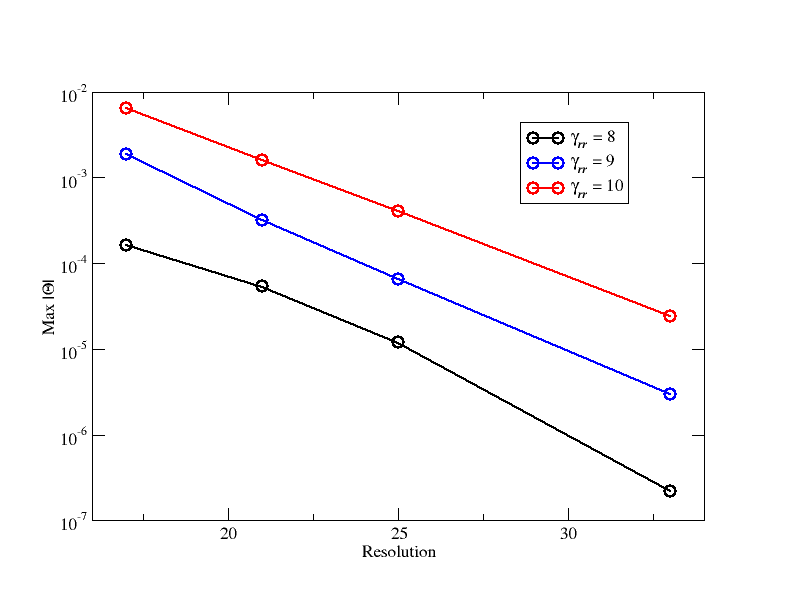}
\caption{
 \label{f:conv_theta} For a non-rotating black hole, maximum value of the expansion $\l|\Theta\r|$ on the horizon, as a function of the resolution (i.e. number of radial collocation points). Three different values of $\gamma_{rr}$ on the horizon are displayed. $\Theta$ clearly goes to zero when precision increases.
}
\end{figure}

Numerical experiments (more details about the numerics can be found in Sec. \ref{s:kerr}) were first constructed in the non-rotating case, that is by setting $\Omega_{\rm BH}=0$. 
If convergence is not impossible to achieve, the numerical system seems to exhibit some instabilities. 
It is observed that the code is much more stable when the boundary condition $\Theta=0$ is relaxed and replaced by the imposition of the value of $\gamma_{rr}$ on the horizon. 
One can then monitor the value of the expansion and check that it is indeed zero. 
This is shown in Fig. \ref{f:conv_theta} where the value of $\Theta$ on the horizon is shown, as a function of the resolution, for three different values of $\gamma_{rr}$ on the horizon. 
When precision increases $\Theta$ goes to zero, for all three different values of $\gamma_{rr}$. 
It shows that all the configurations correspond to valid non-rotating black holes, the different choices of $\gamma_{rr}$ on the horizon corresponding to different masses.

The fact that there is no need to enforce directly $\Theta=0$ in the non-rotating case can actually be understood analytically. 
When $\Omega_{\rm BH}=0$, the only non-vanishing component of the shift is $B^r$ and the spatial line element reduces to ${\rm d}s^2 = A {\rm d}r^2 + B {\rm d} \Omega^2$. 
All the quantities depend only on the radial coordinate. 
Given those expressions, one can compute the expansion and the trace of $K_{ij}$ and find that 

\bea
\label{e:theta_hor_norot}
\Theta &=& \frac{r B B^r \partial_r A + 2 r AB \partial_r B^r + 2 \l(rN\partial B + 2BN\r) \sqrt{A}}{2rABN} \\
\label{e:K_hor_norot}
K &=& \frac{rBB^r \partial_r A + 2r AB \partial_r B^r + 2r AB^r \partial_r B + 4 A B B^r}{2rABN}.
\eea
Inserting $K=0$ in Eq. (\ref{e:theta_hor_norot}) then leads to 
\be
\Theta = \frac{\l(N\sqrt{A} - A B^r\r) \l(2r\partial_r B + 4B\r)}{2rABN},
\ee
which is indeed zero on the horizon given the boundary condition Eq. (\ref{e:lapsehor}). 
So it follows that if the system of equations ensures that $K=0$, then $\Theta=0$ is also guaranteed. 
If the procedure presented in Sec.~\ref{ss:gauge} used to enforce maximal slicing works properly, then there is no need to explicitly impose that $\Theta=0$. 
Given the curves shown in Fig. \ref{f:conv_theta}, it appears to be the case.

When rotation is present (i.e. when $\Omega_{\rm BH}~\not= 0$) the situation is slightly more complicated. 
If one tries the same procedure as in the non-rotating case, one can show that fixing $\gamma_{rr}$ on the horizon does not lead to $\Theta=0$. 
Moreover one can observe that the error does not contain any spherically symmetric component (for instance, if projected onto spherical harmonics, there is no component on $Y_0^0$). 
So in order to maintain some kind of continuity between the non-rotating and rotating cases, one is led to consider the following choice of boundary conditions: 
\begin{itemize}
\item the spherical part of $\gamma_{rr}$ is chosen arbitrarily.
\item the non-spherical part of $\Theta=0$ must be solved.
\end{itemize}

In this paper, spectral methods are used so that the splitting between spherical and non-spherical parts is essentially straightforward. 
Should other numerical methods be used, this may not be as simple but this is beyond the scope of this work.

\subsection{Complete set of boundary conditions \label{ss:final_bc}}

The full set of boundary conditions on the horizon (here a sphere of fixed radius) is given by:

\bea
\label{e:full_set}
N &=& N_0 \l(\theta,\varphi\r) \\
\nonumber
B^i &=& N \tilde{s}^i - \Omega_{\rm BH} m^i \\
\nonumber
{\rm Spherical\, part\,}: \ \gamma_{rr} &=& \gamma_0 \\
\nonumber
{\rm Non-spherical\, part\,}: \ \Theta &=& 0 \\
\nonumber
\gamma_{r\theta} &=& f\l(\theta,\varphi\r) \\
\nonumber
\gamma_{r\varphi} &=& g\l(\theta,\varphi\r)\\
\nonumber
E_{\theta\theta} &=& 0 \\
\nonumber
E_{\theta\varphi} &=& 0 \\
\nonumber
E_{\varphi\varphi} &=& 0.
\eea

In those equations, there are three freely specifiable angular functions $N_0$, $f$ and $g$ and one free number $\gamma_0$. 
A standard choice for those values is: $N=1/2$, $f=g =0$ and $\gamma_0 = 8$. 
Unless otherwise stated, this is what is used throughout this paper.
The equations are solved numerically, using the \textsc{Kadath} library \cite{Grandclement:2009, Kadath}. 
This tool relies on spectral methods to solve systems of partial differential equations and it has been successfully applied to the study of various problems in general relativity and theoretical physics. 

Let us recall that, once the equations are solved, it needs to be checked {\em a posteriori} that the gauge conditions $K=0$ and $V^i=0$ are indeed verified. 
As already stated this is a very important test. 
There is also a need to verify that the spherical part of $\Theta$ vanishes, as it is not explicitly enforced by the set of boundary conditions Eqs. (\ref{e:full_set}).

\section{\label{s:kerr}Kerr black holes}

The simplest and most straightforward application of the equations presented above is the computation of a single rotating black hole in general relativity. 
It must lead to the famous Kerr spacetime \cite{Kerr:1963}. 
However, with the gauge choices used in this paper, the solution is found in coordinates that are not analytical.

The equations solved are (\ref{e:hamilton_ok}-\ref{e:evol_ok}) with the inner boundary conditions (\ref{e:full_set}). 
$\Lambda$ is set to zero and there is no matter.
The system is closed by demanding that flat spacetime is recovered at spatial infinity. 
It simply implies that $N=1$, $B^i=0$ and $\gamma_{ij} = f_{ij}$, where $f_{ij}$ denotes the flat metric. 

For the single black hole problem, the numerical spacetime is decomposed into several (typically 4) spherical shells. 
The last domain extends up to infinity by means of the variable $1/r$ so that boundary conditions are enforced at exact spatial infinity. 
The solutions are found iteratively by means of a Newton-Raphson iteration. 
The first computed configuration is the Schwarzschild one, for which $\Omega_{\rm BH} =0$. 
The angular velocity is then incremented in order to compute a sequence of rotating black holes with different Kerr parameters.

Once a given configuration has been computed, various global, coordinate-independent quantities can be computed. 
The ADM (Arnowitt-Deser-Misner) mass is given by a surface integral at infinity:
\be
\label{e:def_adm}
M_{\rm ADM} = \frac{1}{16 \pi}\int_{r=\infty} f^{ik}  f^{jl}  \l(\bar{D}_j \gamma_{kl} - \bar{D}_k \gamma_{jl}\r) {\rm d}S,
\ee
where $\bar{D}$ denotes the covariant derivative associated with the flat metric $f_{ij}$ and ${\rm d}S$ is the surface element at infinity. 
The spacetime being stationary, one can also define the Komar mass of the system by
\be
\label{e:def_komar}
M_{\rm Komar} =  \frac{1}{4 \pi}\int_{r=\infty} \l(\tilde{s}^i D_i N - K_{ij} \tilde{s}^i \tilde{s}^j \r) {\rm d}S.
\ee
Given Eq.~(\ref{e:bc_shift}), rotation is around the $z$-axis only and the angular momentum is given by
\be
\label{e:def_momentum}
J =  \frac{1}{8 \pi} \int_{r=\infty} K_{ij} m^i \tilde{s}^j {\rm d}S.
\ee

\begin{figure}
    \includegraphics[width=0.48\textwidth,keepaspectratio]{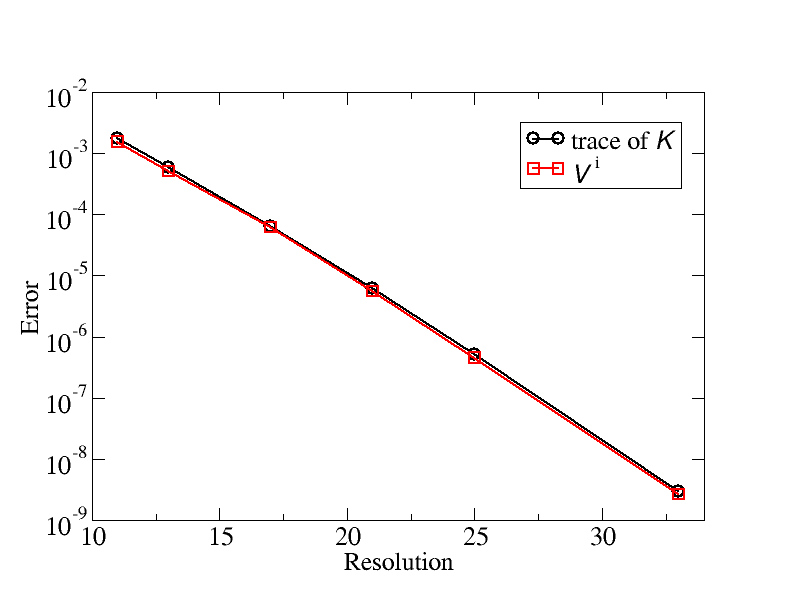}
    \includegraphics[width=0.48\textwidth,keepaspectratio]{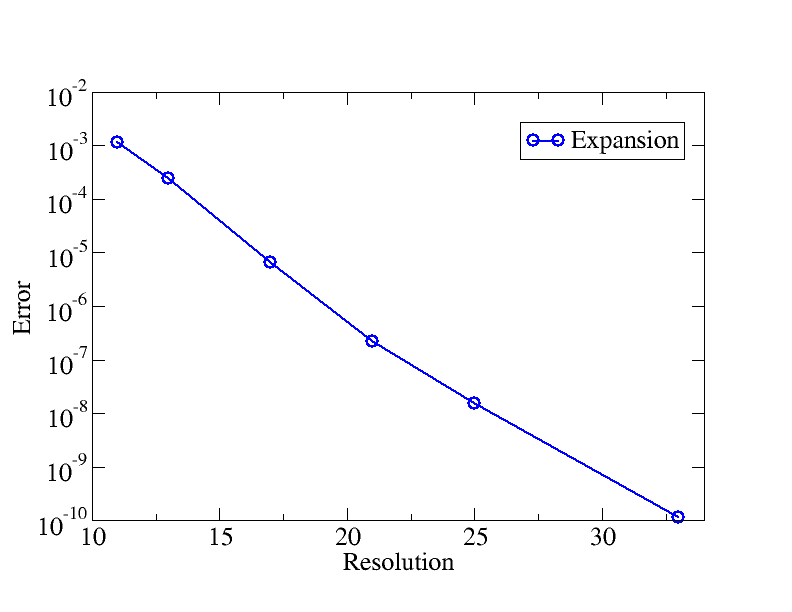} 
     \includegraphics[width=0.48\textwidth,keepaspectratio]{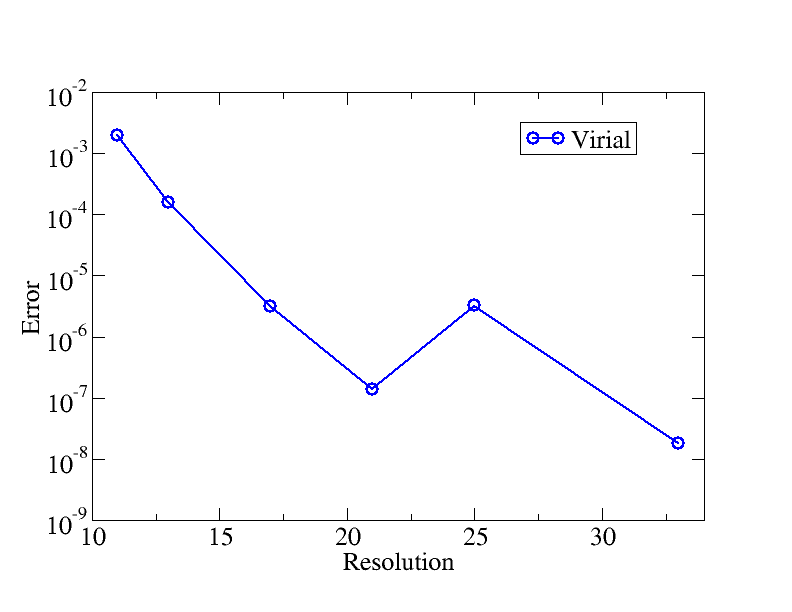}  
      \includegraphics[width=0.48\textwidth,keepaspectratio]{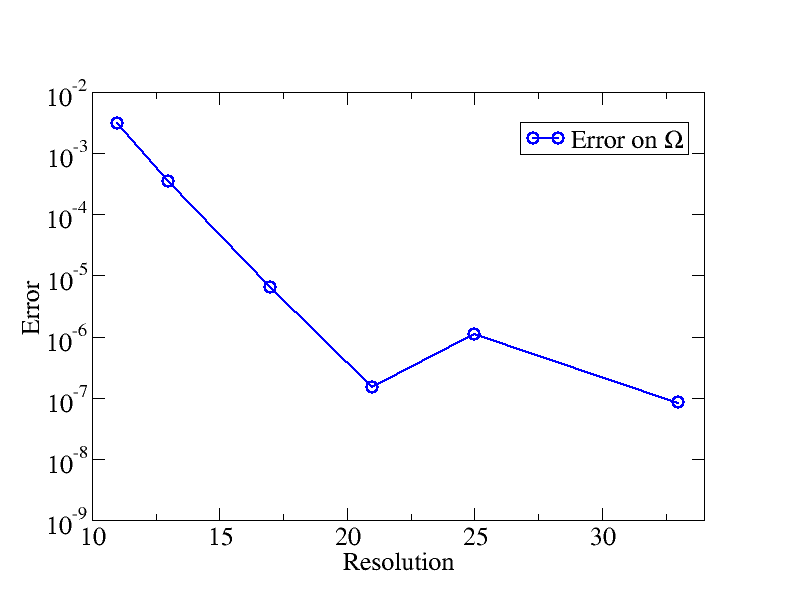}
      
    \caption{\label{f:errors_kerr} Various error indicators for a configuration with $r_{\rm H}=1$ and $\Omega_{\rm BH}=0.1$. All the quantities are shown as a function of the number of spectral coefficients in both the $r$ and $\theta$ directions. The first panel shows the gauge quantities $K$ and $V^i$, the second one the maximum value of $\Theta$ on the horizon, the third one the relative difference between $M_{\rm ADM}$ and $M_{\rm Komar}$ and the last one the relative difference between the numerical and analytical values of $\Omega_{\rm BH}$.}
\end{figure}

Various error indicators are monitored in Fig. \ref{f:errors_kerr}. 
The quantities are plotted as a function of the resolution. 
This corresponds to the number of coefficients of the spectral expansion, with respect to the coordinates $r$ and $\theta$ (in that case the same number is used for both dimensions).
Convergence of the various indicators is shown for a configuration with $r_{\rm H} = 1$ and $\Omega_{\rm BH} = 0.1$, which corresponds to a mass $M_{\rm ADM} \approx 2.18$ and a Kerr parameter $a/M \approx 0.73$. 
The first panel shows that the quantities $K$ and $V^i$ decrease exponentially as resolution increases. 
As already stated, this is an important test that ensures that the system solved coincides with Einstein's equations and that the gauge choices are indeed fulfilled. 
The second panel of Fig. \ref{f:errors_kerr} shows the maximal value of $\Theta$ (Eq. \ref{e:expansion}) on the horizon. 
As seen in Sec. \ref{ss:expansion} the spherical part of $\Theta=0$ is not solved numerically and this curve shows that it is indeed verified as it goes to zero exponentially.

The last two panels involve the computation of global quantities and so the errors are slightly bigger. 
Indeed the computation of those quantities involve surface integrals at infinity that introduce additional numerical errors, when compared to quantities like $K$ or $V^i$. 
This can explain the fact that convergence is less regular, as seen with the values for a resolution of 25. 
The saturation level is also somewhat higher, with a value of about $10^{-7~\, -8}$. 
This is not surprising as the Newton-Raphson algorithm was stopped at a threshold of $10^{-8}$. 
That being said, the last two panels of Fig. \ref{f:errors_kerr} still show a good convergence with resolution. 
The third one  shows the relative difference between the ADM and Komar masses, as it is known that the two must be equal in that case (see Sec. 8.6.2 of \cite{Gourgoulhon:2007} for more details on this equality). 
The last panel shows the relative difference from the numerical angular velocity $\Omega_{\rm BH}$ and the one computed from $M_{\rm ADM}$ and $a$ by the analytic expression 
\be
\label{e:omega_ana}
\Omega = \frac{a}{2M\l(M+\sqrt{M^2-a^2}\r)}.
\ee

\begin{figure}
    \includegraphics[width=0.48\textwidth,keepaspectratio]{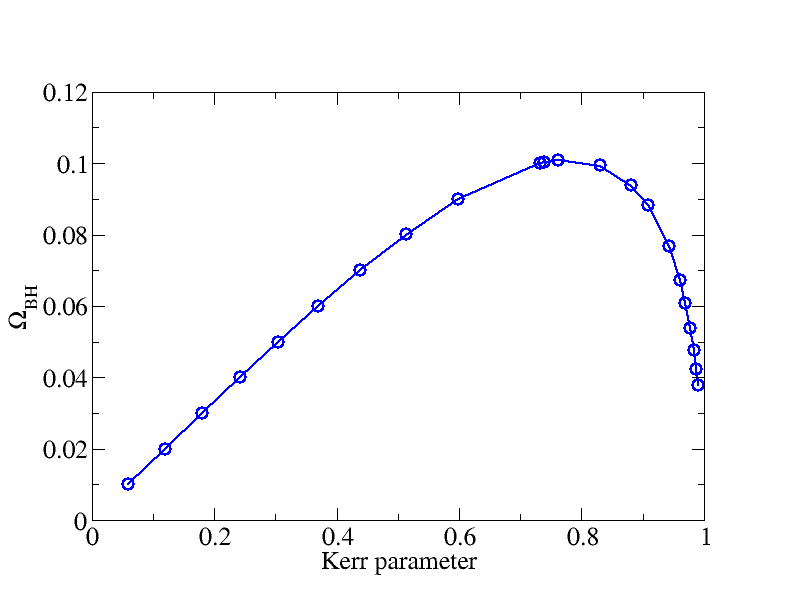}
     \includegraphics[width=0.48\textwidth,keepaspectratio]{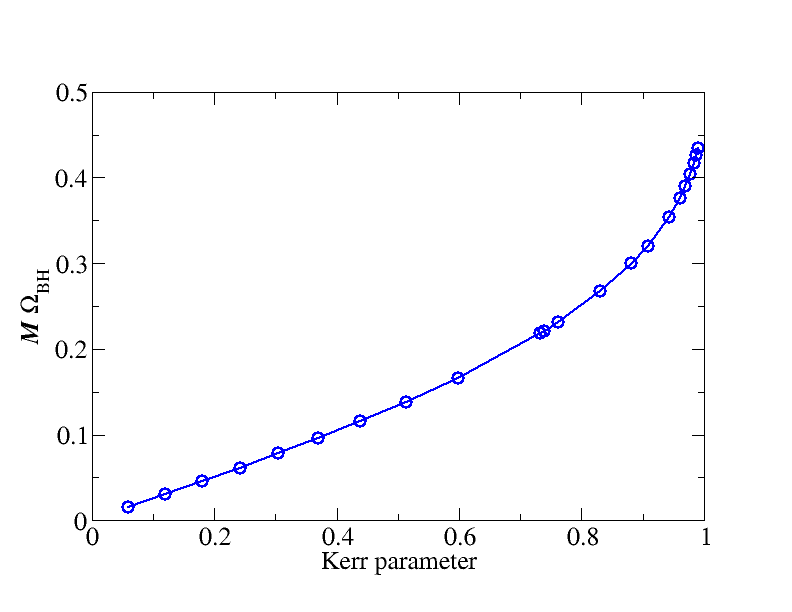}
    \caption{\label{f:sequence_kerr} The first panel shows $\Omega_{\rm BH}$ as a function of $a/M$ for a sequence of constant radius $r_{\rm H} = 1$ and the second $M\Omega_{\rm BH}$ as a function of $a/M$ also. Configurations are computed with the highest resolution at hand.}
\end{figure}

Using the highest resolution at hand (33 points in $r$ and $\theta$), one can compute a sequence of Kerr black holes, with different values of the Kerr parameter $a/M$. 
It is well known that this parameter goes from 0 (Schwarzschild black hole) to 1 (extremal Kerr black hole). 
In order to compute the sequence, one can start from the non-rotating solution and increase $\Omega_{\rm BH}$ step-by-step. 
If this does work properly at first (i.e. for moderate values of $a/M$), a technical difficulty stems from the fact that $\Omega_{\rm BH}$ is not a monotonic function of the Kerr parameter, when the radius $r_{\rm H}$ is fixed. 
This is clearly seen in the first panel of Fig. \ref{f:sequence_kerr}. 
In order to be able to pass the maximum, the value of $\Omega_{\rm BH}$ can be made an unknown of the numerical problem and the condition that $a/M$ has a given value can be added to the system. 
The \textsc{Kadath} library enables the use of such global unknowns (i.e. unknowns that are not fields but numbers). 
Doing so one can reach high values of the Kerr parameters, that correspond to small values of $\Omega_{\rm BH}$. 
In the second panel the angular velocity is also shown, as a function of $a/M$, but this time scaled with the ADM mass. 
The fact that the mass is not constant along the sequence (it is $r_{\rm H}$ that is fixed), explains the different behavior between the two panels. 
As is expected (see Eq. (\ref{e:omega_ana})), $M \Omega_{\rm BH}$ goes to $1/2$ when one gets closer to the extremal case. 
Let us finally mention that very high values of $a/M \approx 0.99$ can be reached without much trouble and while maintaining an accuracy of about $10^{-7}$. 
If need be, even higher values could be computed.

\begin{figure}
    \includegraphics[width=0.3\textwidth,keepaspectratio]{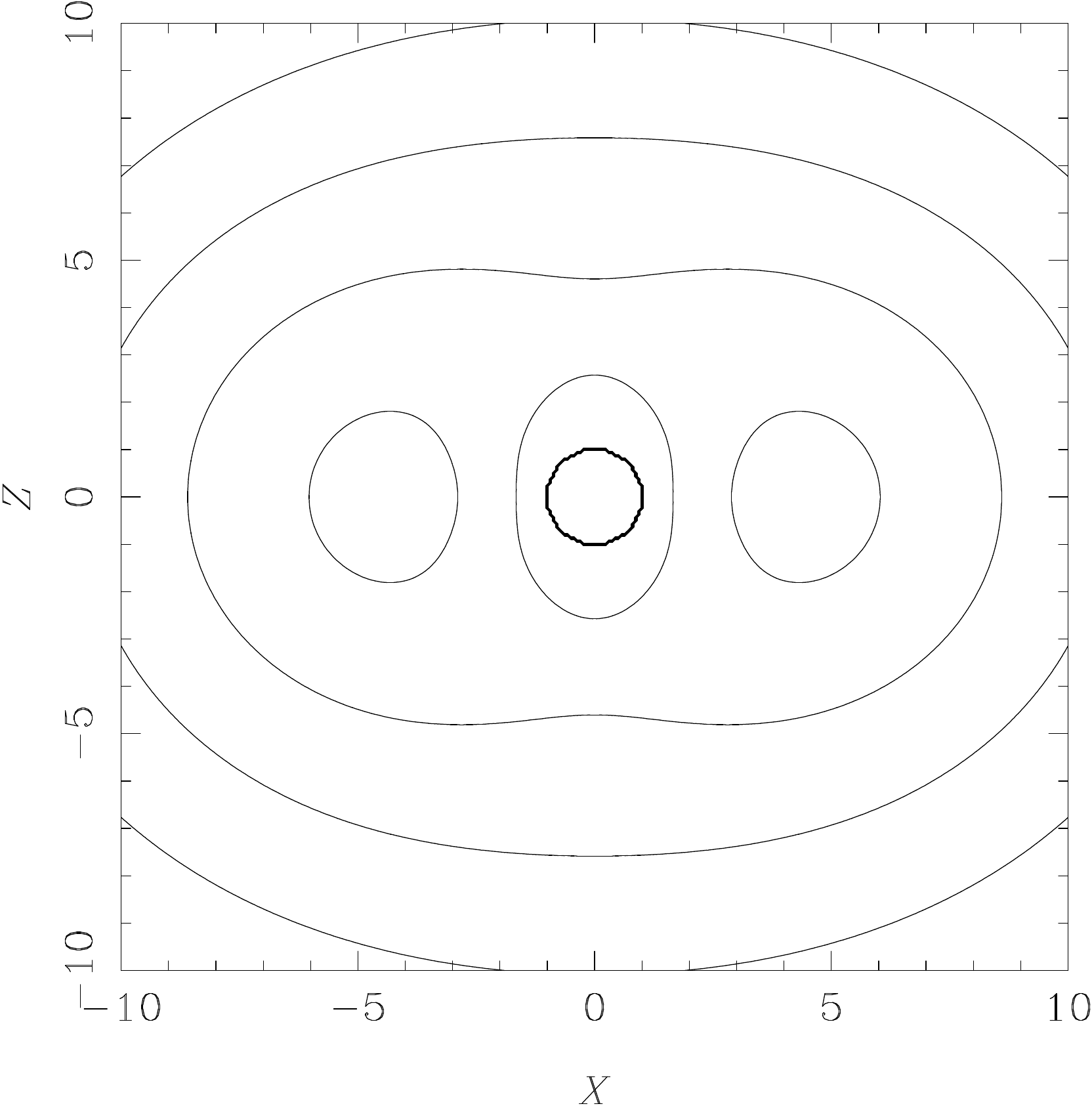}
    \includegraphics[width=0.3\textwidth,keepaspectratio]{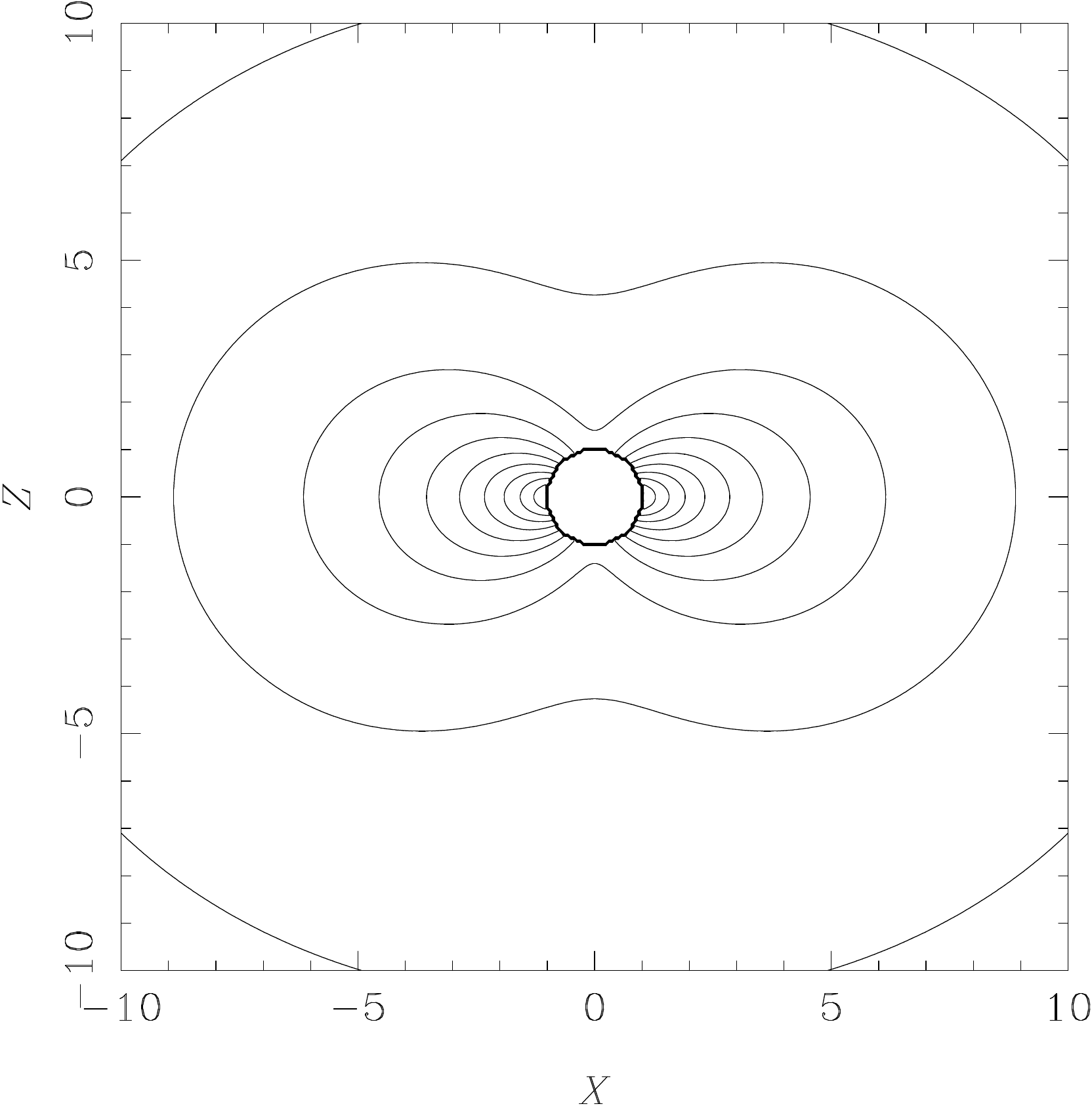}
     \includegraphics[width=0.3\textwidth,keepaspectratio]{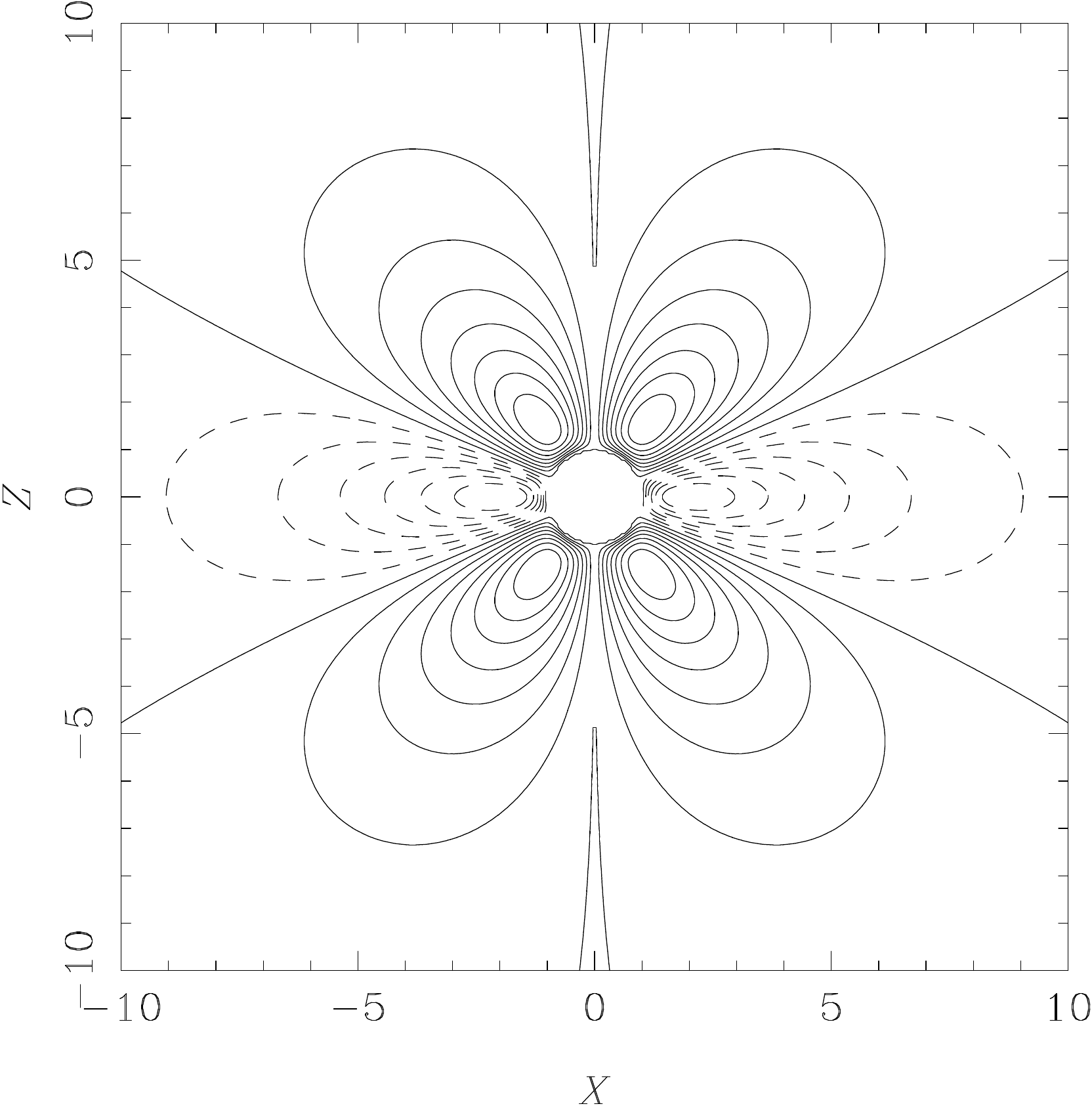}
    \caption{\label{f:profs_kerr} Contours of $N$ (first panel), $B^r$ (second panel) and $\gamma_{r\varphi}$ (third panel), in the $xz$-plane, for the configuration with $a/M \approx 0.99$.} 
\end{figure}

As an illustration, in Fig. \ref{f:profs_kerr}, various contours of some fields are shown, in the $xz$-plane, for the configuration with the highest value of $a/M \approx 0.99$.

\section{\label{s:MTZ} MTZ black holes}

In this section the formalism presented in Sec. \ref{s:equations} is applied to a class of non-rotating black holes with a negative cosmological constant and a minimally coupled scalar field. 
This solution was obtained analytically by Martínez, Troncoso and Zanelli (MTZ) in \cite{Martinez:2004}.

\subsection{\label{ss:MTZ_system} Analytic solution and adjustments to the system of equations}

The field equations are the Einstein ones with a negative cosmological constant $\Lambda = -3/\ell_{\rm ADS}^2$, as well as the Klein-Gordon equation for a real  scalar field $\phi$, minimally coupled to gravity.

The stress-energy tensor is given by 
\be \label{e:MTZ_Tmunu}
T_{\mu\nu}=\nabla_\mu\phi \nabla_\nu\phi - \dfrac{1}{2}g_{\mu\nu}g^{\alpha\beta}\nabla_\alpha\phi \nabla_\beta\phi - g_{\mu\nu}V(\phi) ,
\ee
and the interaction potential is
\be \label{e:MTZ_potential}
V(\phi) = -\dfrac{3}{4\pi \ell_{\rm ADS}^2}\sinh^2{\sqrt{\dfrac{4\pi}{3}}\phi}.
\ee

The 3+1 matter terms in that particular case are given in Appendix \ref{ss:MTZ_3+1}.

In \cite{Martinez:2004}, a family of black hole solutions parameterized by an integration constant $\mu$ (linked to the black hole mass) is found :
\be \label{e:MTZ_analytic_metric}
\mathrm{d}s^2 = \dfrac{r(r+2\mu)}{(r+\mu)^2} \l[-\l( \dfrac{r^2}{\ell_{\rm ADS}^2} - \l(1+\dfrac{\mu}{r}\r)^2\r)\mathrm{d}t^2 + \l( \dfrac{r^2}{\ell_{\rm ADS}^2} - \l(1+\dfrac{\mu}{r}\r)^2\r)^{-1} \mathrm{d}r^2 + r^2 \mathrm{d}\sigma^2 \r]
\ee
\be \label{e:MTZ_analytic_phi}
\phi = \sqrt{\dfrac{3}{4\pi}}\arctanh \l( \dfrac{\mu}{r+\mu}\r) .
\ee

The term $\mathrm{d}\sigma^2$ represents the line element of a two-dimensional manifold with constant negative curvature. In this work, it is chosen as:
\be
    \label{e:MTZ_hyperbolic_metric}
    \mathrm{d}\sigma^2 = \dfrac{1}{\cos^2\theta}(\mathrm{d}\theta^2 + \sin^2\theta ~\mathrm{d}\varphi^2).
\ee

This is to be contrasted with  \cite{Martinez:2004} where this two-dimensional manifold (and hence the horizon) is assumed to be compact. 
The form (\ref{e:MTZ_hyperbolic_metric}) corresponds to coordinates on the hyperbolic plane.

Furthermore, the vector field $(\partial_\varphi)^i$ appearing in the boundary conditions (\ref{e:full_set}) is not a conformal Killing field of the surface described by (\ref{e:MTZ_hyperbolic_metric}). 
However, as the solution represents a non-rotating black hole, for which $\Omega_{\rm BH}=0$, this is not an issue.
 
For $r\xrightarrow{}\infty$ or equivalently $\mu=0$, anti-de Sitter spacetime with constant negative (4D) curvature $^4R=2\Lambda$ is recovered. The solution is said to be asymptotically anti-de Sitter (AADS).

The constant $\mu$ is bounded from below by $\mu > \dfrac{\ell_{\rm ADS}}{4}$ and  $r$ must be larger than the maximum of 0 and $-2\mu$.
Under those conditions, there is a horizon located at $r = r_+$ with 
\be
    \label{e:MTZ_r+}
    r_+ = \dfrac{\ell_{\rm ADS}}{2}\l( 1+\sqrt{1+\dfrac{4\mu}{\ell_{\rm ADS}}} \r).
\ee
The corresponding apparent horizon is a surface of constant negative curvature.

In order to describe surfaces of negative curvature, one considers the following reference spatial metric:
\be
\gamma_{ij}^{\rm ref} \mathrm{d}x^i \mathrm{d}x^j = \mathrm{d}r^2+r^2\mathrm{d}\sigma^2 = \mathrm{d}r^2 + \dfrac{r^2}{\cos^2\theta}(\mathrm{d}\theta^2+\sin^2\theta \mathrm{d}\varphi ^2).
\ee
This reference metric is used when defining the Christoffel symbols of the real metric. 
More precisely, what is computed numerically is the difference between the Christoffel symbols of $\gamma_{ij}$ and those of $\gamma_{ij}^{\rm ref}$.
This differs from more usual situations where the reference metric is the flat one. The numerical description of the reference metric is done via the basis:
\be
{\bm e_r} = {\bm \partial_r} \quad ; \quad {\bm e_\theta} = \frac{\cos\theta}{r}{\bm \partial_\theta} \quad ; \quad {\bm e_\varphi} = \frac{\cos\theta}{r\sin\theta} {\bm \partial_\varphi},
\ee
which makes its orthonormal. 

As the solutions are not asymptotically flat but AADS, the ADS boundary must be dealt with carefully, as some metric quantities diverge at this location. 
In order to do so, a particular form of the ADS spacetime is chosen 
\be
\label{e:MTZ_ADS_metric}
{\rm d} s^2 = - \frac{\cos^2 \l(\log \frac{R_{\rm ADS}}{r}\r)}{\sin^2 \l(\log \frac{R_{\rm ADS}}{r}\r)} {\rm d}t^2 + 
\frac{\ell_{\rm ADS}^2}{r^2 \sin^2 \l(\log \frac{R_{\rm ADS}}{r}\r)} \gamma_{ij}^{\rm ref} {\rm d}x^i {\rm d}x^j.
\ee

$R_{\rm ADS}$ is a freely specifiable value that gives the position of the boundary of the ADS spacetime, which is located at $r=R_{\rm ADS}$. 
The form (\ref{e:MTZ_ADS_metric}) is chosen because the spatial metric relates conformally to the reference one 
\be
\gamma_{ij}^{\rm ADS} = \frac{\ell_{\rm ADS}^2}{r^2 \sin^2 \l(\log \frac{R_{\rm ADS}}{r}\r)} \gamma_{ij}^{\rm ref}.
\ee
The ADS spatial metric is the one which is used in defining the spatial harmonic gauge, meaning that $\bar{\gamma}_{ij} = \gamma_{ij}^{\rm ADS}$ (see Eq. (\ref{e:harmonic})).

From Eq. (\ref{e:MTZ_ADS_metric}) one can see that the lapse and spatial metric diverge at this boundary. 
In order to allow for a numerical treatment there is a need to regularize the divergences. 
This is done by defining the conformal factor

\be
\label{e:MTZ_Omega}
\Omega \equiv \sin\l(\log\dfrac{R_{\rm ADS}}{r}\r).
\ee

$\Omega$ can then be used to define the following regularized quantities (denoted by a tilde):

\bea
\label{e:MTZ_reg_unknowns}
   \Tilde{N} & \equiv & \Omega N \\
   \Tilde{B}^i & \equiv & B^i \\
   \Tilde{\gamma}_{ij} & \equiv & \Omega^2\gamma_{ij} \\
   \Tilde{\phi} & \equiv & \phi. 
\eea

The outer boundary conditions, where $r = R_{\rm ADS}$, are then $\tilde{N} = 1$, $\tilde{B}^i = 0$, $\tilde{\gamma}_{ij} = \displaystyle\frac{\ell^2_{\rm ADS}}{R^2_{\rm ADS}} \gamma_{ij}^{\rm ref}$ and $\tilde{\phi}=0$.

All the quantities appearing in the equations must be regularized near the ADS boundary, along with the equations themselves (see Appendix \ref{ss:MTZ_reg} for explicit expressions). 

This regularization procedure applies in the outermost numerical domain, which is a spherical shell extending up to the ADS boundary at $r=R_{\rm ADS}$. 
Proper continuity of the fields and their radial derivatives across the boundary with the inner domain is enforced. 
This translates into the following non trivial conditions (where superscripts $(I)$ and $(O)$ stand for inner and outer respectively):
\bea
    \label{e:MTZ_reg_matching_lapse}
    \partial_r\Tilde{N}^{(O)} & = & N^{(I)}\partial_r\Omega + \Omega\partial_r N^{(I)}, \\
    \label{e:MTZ_reg_matching_gamma}
    \partial_r\Tilde{\gamma}_{ij}^{(O)} & = & \Omega^2 \partial_r\gamma_{ij}^{(I)} + 2\Omega(\partial_r\Omega) \gamma_{ij}^{(I)}.
\eea

Last but not least, it must be noted that the the Klein-Gordon equation is degenerate on the horizon. 
From its expression, given by Eq. (\ref{e:3+1_KG}), one can notice that the only contribution to the principal symbol for double $r$ partial derivatives is $\l(\gamma^{rr}-\dfrac{(B^r)^2}{N^2} \r)\partial^2_{rr}\phi$. 
It is the same factor that appears in the Einstein evolution equations (see Sec. \ref{ss:dege}) and it vanishes on the horizon with the boundary condition (\ref{e:lapsehor}). 
Therefore, the Klein-Gordon equation is degenerate on the horizon and there is no need to impose any additional boundary condition on the scalar field itself.

\subsection{\label{ss:MTZ_results} Results}

The cosmological constant is chosen so that $\ell_{\rm ADS}=20$. 
Space is split into two spherical shells. 
The outer one goes from $r=30$ to $r=R_{ADS}=40$ and so extends up to the ADS boundary. 
The radius of the horizon is varied in order to compute different configurations.

At the inner boundary, which is the apparent horizon, one sets $N_0 = 1/2$, $\gamma_0 = 1$ and $f=g=0$ (see Eqs. (\ref{e:full_set})). 
Recall that the MTZ solution is a static one so that $\Omega_{\rm BH}=0$.

\begin{figure}  
   \includegraphics[width=0.48\textwidth]{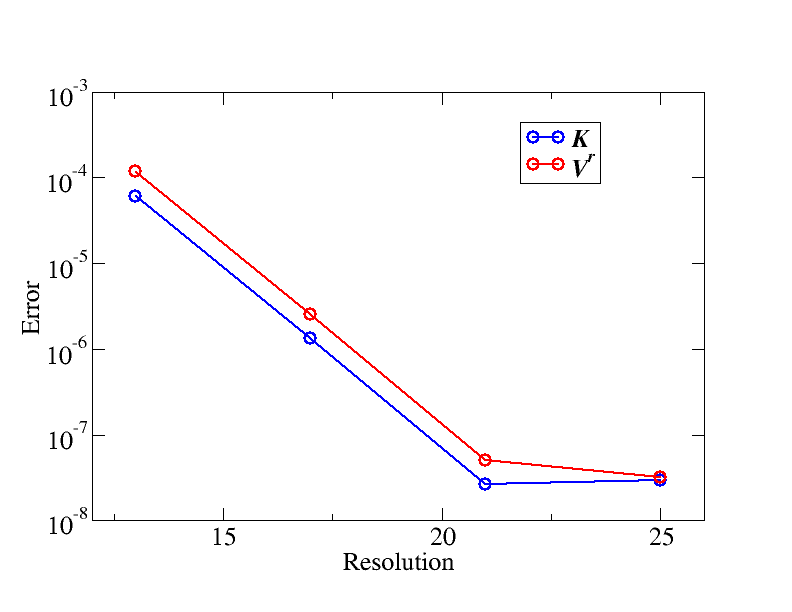}
   \includegraphics[width=0.48\textwidth]{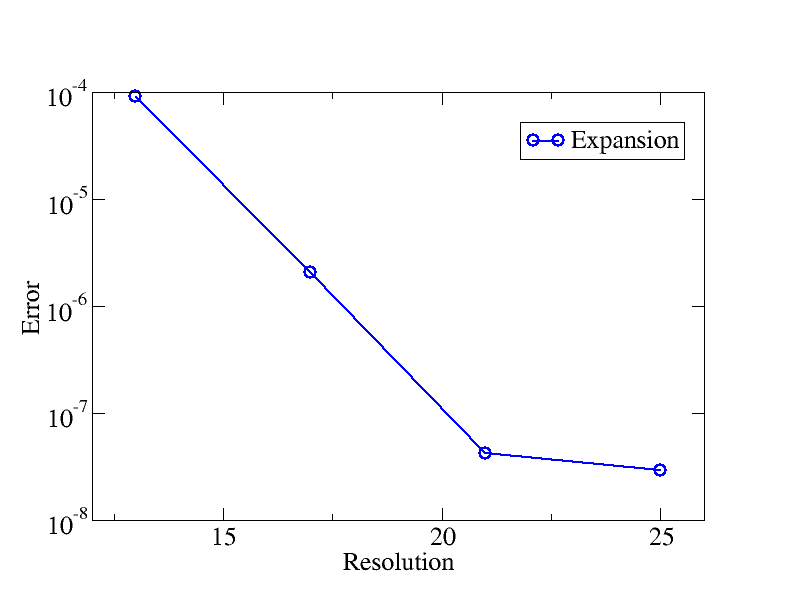}
    \caption{\label{f:MTZ_conv} Maximum value of the gauge quantities $K$ and $V^r$ and of the expansion on the horizon, as a function of the radial number of coefficients. Spectral convergence is clearly seen.}
\end{figure}

As in the Kerr black hole case, spectral convergence of the gauge quantities and of the expansion on the horizon is monitored.
Figure \ref{f:MTZ_conv} shows these quantities as a function of the radial resolution and convergence is clear. 
The saturation seen at $10^{-8}$ is due to the Newton-Raphson solver that is stopped at this level.

\begin{figure} 
  \includegraphics[width=0.48\textwidth]{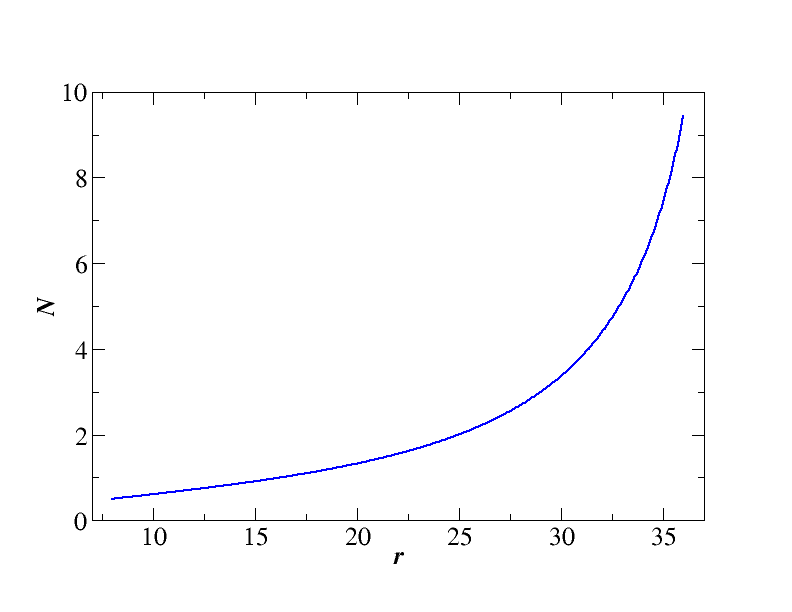}
  \includegraphics[width=0.48\textwidth]{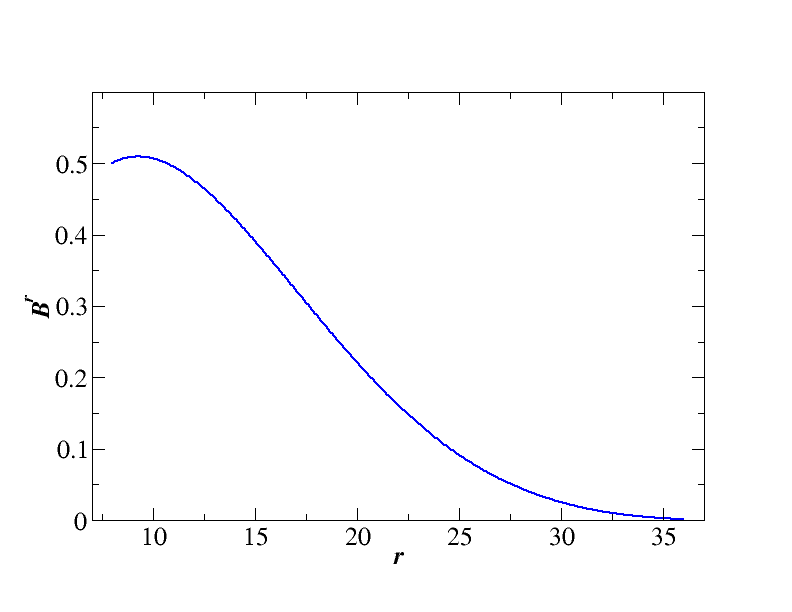}
   \includegraphics[width=0.48\textwidth]{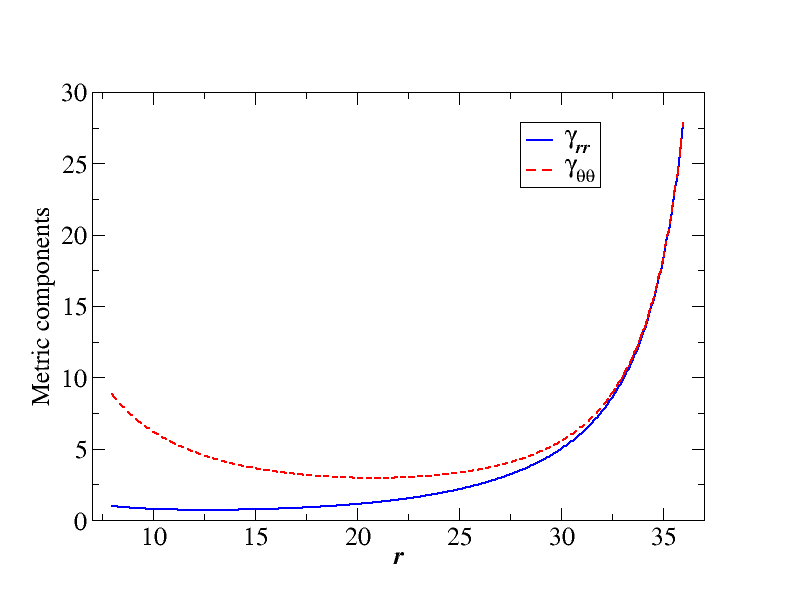}
    \includegraphics[width=0.48\textwidth]{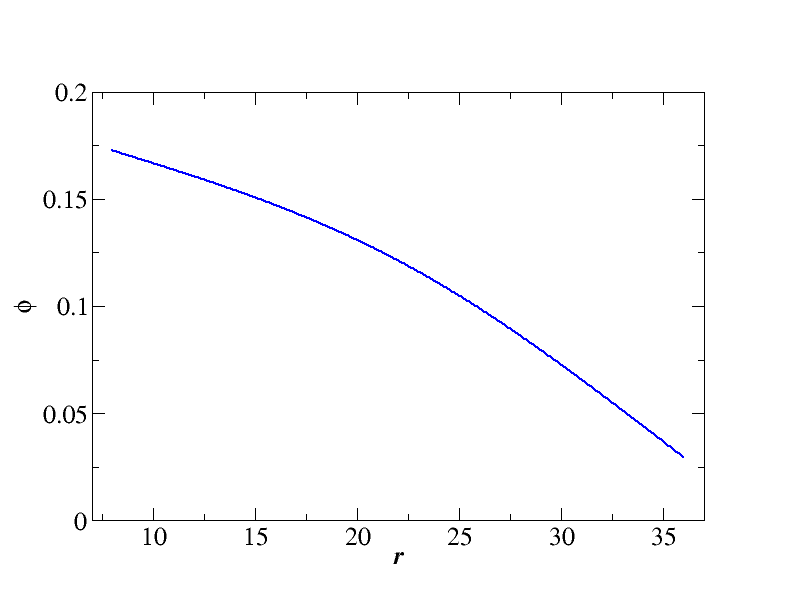}
   \caption{\label{fig:MTZ_profs} Profiles of various fields (first panel $N$, second $B^r$, third $\gamma_{rr}$ and $\gamma_{\theta\theta}$ and fourth $\phi$) for $r_{\mathrm{H}}=8$ and 21 radial coefficients in each domains. As expected, the lapse and spatial metric diverge near the ADS border.}
\end{figure}

Profiles of various fields are shown in Fig. \ref{fig:MTZ_profs} for a configuration with $r_{\rm H} =8$. 
Divergence of the lapse and of the spatial metric at the ADS boundary is noticeable (remember that this boundary is at $R_{\rm ADS} = 40$). 
The smoothness of the curves of $N$ and $\gamma_{ij}$ illustrates the good behavior of the regularization procedure..

In order to compare the numerical results with the solution (\ref{e:MTZ_analytic_metric}), coordinate-independent quantities must be compared.
The four-dimensional Ricci scalar on the horizon $^4R_\mathrm{H}(\phi_\mathrm{H})$ is one possibility.
It can be computed in the 3+1 formalism by Eq. (3.75) \cite{Gourgoulhon:2007}, written here in the case $K=0$ 
\be
    \label{e:MTZ_4D_Ricci}
    ^4R = R + K_{ij}K^{ij} - \dfrac{2}{N}D_iD^iN.
\ee

\begin{figure}
    \centering
    \includegraphics[width=0.8\linewidth]{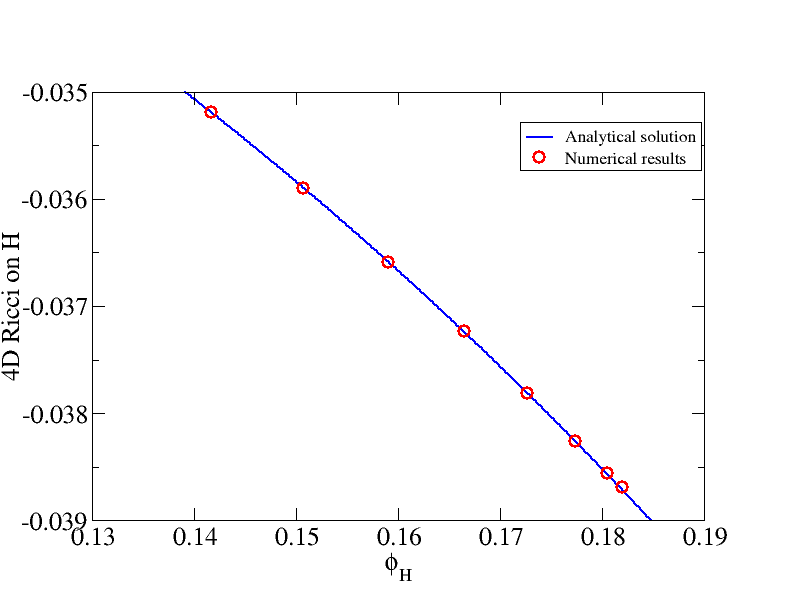}
    \caption{\label{fig:MTZ_R4D} Four-dimensional Ricci scalar on the horizon as a function of the value of the scalar field on the horizon. The solid line is computed from the analytic solution, while the circles correspond to the numerical results. Horizon radii $r_\mathrm{H}$ range from 6.5 to 10 (from right to left). Large radii correspond to reduced scalar field and weaker curvature.
    }
\end{figure}

A sequence of MTZ black holes with various radii for the horizon, ranging between $r_\mathrm{H}=6.5$ and $r_\mathrm{H}=10$ is computed.
Although each solution cannot directly be linked to a specific value of $\mu$, the sequence consists of black holes with different masses. 
The four-dimensional Ricci scalar on the horizon, as a function of the value of the scalar field on the horizon is shown in Fig. \ref{fig:MTZ_R4D}.
The circles denote the numerical results and the solid curve the analytical one. 
A very good agreement between the two is achieved.
The smaller the horizon radius, the more intense the scalar field on the horizon and the larger the absolute value of the curvature on the horizon.
This means that small radii correspond to more relativistic configurations.

\section{\label{s:hairy} Black holes with scalar hairs}

In this section, the construction of Kerr black holes with scalar hairs is explained. 
This is the same system as computed by \cite{Herdeiro:2014, Herdeiro:2015} but using a different choice of coordinates.

\subsection{\label{ss:hairy_clouds} Equations for the scalar field}

The class of hairy black holes constructed here relies on the existence of a complex scalar field $\Phi$. 
This field is minimally coupled to gravity and solutions containing a horizon can be found. 
The scalar field obeys the Klein-Gordon equation $\nabla_\mu \nabla^\mu \Phi = \mu^2 \Phi$ where $\nabla$ denotes the covariant derivative of the four-dimensional metric. 
The right-hand-side corresponds to the case of a free massive field which has a potential $V\l(\l|\Phi\r|^2\r) = \mu^2 \l|\Phi\r|^2$, where $\mu$ is the mass of the field.

In previous works the field is assumed to have the form $\Phi = \phi\l(r, \theta\r) \exp \l[i\l(\omega t- k \varphi\r)\r]$, where $\omega$ is the angular velocity and $k$ an integer dubbed the rotational quantum number. 
Given the expression of the action (see \cite{Herdeiro:2014, Herdeiro:2015}), which has a $U\l(1\r)$ symmetry, the resulting spacetimes are axisymmetric (the two Killing vectors being $\l(\partial_t\r)^\mu$ and $\l(\partial_\varphi\r)^\mu$) and the quantities $\omega$ and $k$ appear as parameters of the solutions. 
When the ansatz is inserted into the Klein-Gordon equation, it leads to an expression of the form
\be
\l(R_{\rm KG} + i I_{\rm KG} \r) \exp\l[i\l(\omega t- k \varphi\r)\r] = 0.
\ee
In \cite {Herdeiro:2014, Herdeiro:2015} the part $R_{\rm KG}$ is the only one considered and it is the equation fixing the value of the amplitude $\phi$. 
Indeed, given the coordinates used, and in particular the fact that the only non-vanishing component of the shift is $B^\varphi$, it is easy to show that $I_{\rm KG}$ is identically zero. 
With the coordinate system introduced in this work, this is no longer the case (basically because the shift has a non-vanishing radial component). 
It means that the original ansatz does not pass through the equation and that a more general form must be used.

One considers the following form for the scalar field, introducing an additional imaginary component
\be
\label{e:scalar_field}
\Phi = \l(R_\Phi \l(r,\theta \r)  + i I_\Phi\l(r,\theta \r)  \r)  \exp\l[i\l(\omega t- k \varphi\r)\r] = 0.
\ee

Inserting (\ref{e:scalar_field}) in the Klein-Gordon equation and using the 3+1 decomposition of spacetime, one can find the expressions for $R_{\rm KG}$ and $I_{\rm KG}$ in terms of $R_\Phi$ and $I_\Phi$. One gets

\bea
\label{e:RKG}
R_{\rm KG} &=& \frac{\l(\omega + k B^i \delta_i^\varphi\r)^2}{N^2} R_\Phi + \frac{1}{N} D_i \l(N\l(\gamma^{ij} - \frac{B^iB^j}{N^2}\r)\r) D_j R_\Phi + 
\l(\gamma^{ij} - \frac{B^iB^j}{N^2}\r) D_i D_j R_\Phi\\
\nonumber
&& - \gamma^{ij} k^2 \delta_i^\varphi \delta_j^\varphi R_\Phi  - \mu^2 R_\Phi \\
\nonumber
&& - \omega \frac{B^i}{N^2} D_i I_\Phi - \frac{\omega}{N} D_i\l(\frac{B^i}{N}I_\Phi\r) 
+ \frac{1}{N} D_i \l(N\l(\gamma^{ij} - \frac{B^iB^j}{N^2}\r) k \delta_j^\varphi I_\Phi\r) + \l(\gamma^{ij} - \frac{B^iB^j}{N^2}\r) k \delta_i^\varphi D_j I_\Phi,
\eea
where $\delta_i^\varphi$ is the gradient of $\varphi$. 
Expressed in the spherical orthonormal basis used in this paper, it is given by $\delta_i^\varphi = \l(0, 0, \frac{1}{r \sin \theta}\r)$. 
Divisions by $\sin\theta$ -- which vanishes on the $z$-axis -- are performed using the coefficients of the spectral expansion of the fields, in order to avoid divergences. 

Eq. (\ref{e:RKG}) is of the form $A \l(R_\Phi\r) + B \l(I_\Phi\r) = 0 $ where $A$ and $B$ are linear differential operators of the scalar field ($A$ is second order and $B$ first order). 
Not surprisingly, given the ansatz used, the expression of $I_{\rm KG}$ is very similar to the one for $R_{\rm KG}$. 
One finds that $I_{\rm KG} = A \l(I_\Phi\r) - B\l(R_\Phi\r)$.

As a first step, the metric fields are fixed and only the Klein-Gordon equation is solved. 
It means that the backreaction of the field on the metric is neglected. This is known as a cloud solution. 
For the metric fields one uses the Kerr black hole configurations computed in Sec. \ref{s:kerr}. 
Let us mention that, following \cite{Herdeiro:2014, Herdeiro:2015}, the angular velocity of the black hole and of the field are linked by $\omega = k \Omega_{\rm BH}$, a condition that prevents the field from having a flux across the horizon. 
This condition is enforced throughout this paper.

The resolution of this linear problem needs to be dealt with carefully. 
First one can notice that the part of $R_{\rm KG}$ that contains second order radial derivatives is $\l(\gamma^{rr} - \frac{B^rB^r}{N^2}\r) \partial_r^2 R_\Phi$. 
As already seen in Sec. \ref{s:equations} and \ref{s:MTZ}, with the boundary conditions used, this term vanishes on the horizon. 
The same is true for $I_{\rm KG}$, with respect to $I_\Phi$. 
It means that those equations do not require any inner boundary condition (see Sec. \ref{ss:dege}). 
So there is no need to enforce anything on the fields at the horizon.

Two other properties of the linear Klein-Gordon equations must be taken into account. 
First notice that $R_\Phi = I_\Phi = 0$ is a solution so that one needs to prevent the code from converging to this trivial configuration. 
Also, it is expected that solutions can be found only for a discrete set of the physical parameters ($k$, $\Omega_{\rm BH}$, $\mu$ in particular). 
The numerical procedure should be able to find those values. 
Those two features are typical of linear systems and have been dealt with successfully in the context of spacetimes with cosmological constants in \cite{Fodor:2013, Fodor:2015} and the reader should refer to those publications for details about the procedure. 
It is only briefly sketched below. 

\begin{figure}
\includegraphics[width=.7\textwidth,keepaspectratio]{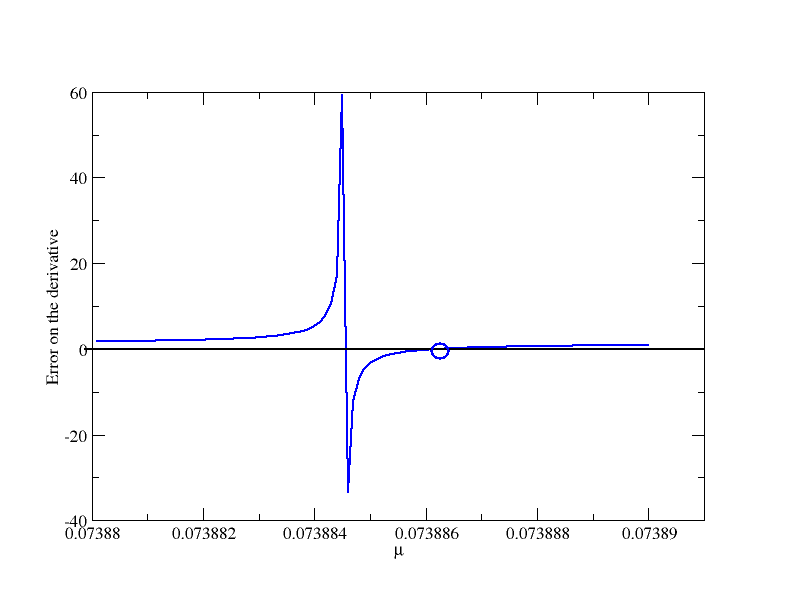}
\caption{\label{f:scan_der} Discontinuity of the radial derivative of $R_\Phi$, as a function of $\mu$. The circle denotes the location of the true solution for which $\Phi$ is regular. The background corresponds to a fastly rotating Kerr black hole with $M \approx 5. $ and $a/M \approx 0.95$.}
\end{figure}

The equations being second order, they should be solved by demanding the matching of the fields and their normal derivative across the boundaries of the various numerical domains. 
For one such boundary, the continuity of one angular spectral coefficient is relaxed and replaced by the condition that this coefficient has a given value ($1$ for instance). 
By construction, this prevents the code from going to the trivial solution. 
The value of the coefficient itself is unimportant, the problem being linear. 
However, in general, that would lead to a solution with a discontinuity on the derivative. 
The next step is then to scan the possible values of the parameter space and it appears that for some values, the error on the discontinuity vanishes. 
An example of that is shown in Fig. \ref{f:scan_der} where the error on the derivative of $R_\Phi$ is plotted, as a function of the parameter $\mu$, around the value for a true solution. 
The correct value of $\mu$ can be determined by a dichotomy algorithm. 
However, a good precision is required, as the error varies very fast with $\mu$ (notice the abscissa range in Fig. \ref{f:scan_der}). 
As observed in \cite{Fodor:2013, Fodor:2015}, there are several possible values of the parameters that are admissible. 
They correspond to different number of nodes of the amplitude $\l|\Phi\r|$. 
Only nodeless configurations are considered here.

An additional difficulty arises from the fact that the quantities $R_\Phi$ and $I_\Phi$ are determined up to a constant phase ; the fields defined as $R'_\Phi = R_\Phi \cos\alpha  - I_\Phi \sin\alpha $ and $I'_\Phi = R_\Phi \sin\alpha  + I_\Phi \cos\alpha $ are also a valid solution of the system. 
The value of this phase must be enforced when solving numerically the equations, otherwise the Newton-Raphson iteration would fail. 
To do so, one follows that same procedure for $I_\Phi$ as for $R_\Phi$: the continuity of the derivative is relaxed and replaced by a condition on the value of one coefficient of $I_\Phi$, at an arbitrary location. 
It appears that different values of $I_\Phi$ all lead to valid solutions that correspond to different choices of $\alpha$. 
As this technique is also implemented in the full system case (i.e. the one with gravitation), more details are given about this in Sec. \ref{ss:hairy_results}.

The cloud solutions described in this section are only used as an initial guess to get the configurations of the coupled Einstein-Klein-Gordon system so that they are not investigated much here. 
However, their validity has been carefully checked by comparing with results coming from the direct resolution of the Teukolsky equation as done in \cite{Garcia:2018} for instance.

\subsection{\label{ss:hairy_full} Coupling with gravity}

When the scalar field is coupled to gravity, one must take into account the stress-energy tensor of the field, which reads 
$T_{\mu\nu} = \nabla_{\l(\mu\r.} \Phi \nabla_{\l.\nu\r)} \bar{\Phi} - \frac{1}{2} g_{\mu\nu} \l(\nabla_\mu \Phi \nabla^\mu \bar{\Phi} +V\l(\l|\Phi\r|^2\r)\r)$. 
Using the ansatz of the field (\ref{e:scalar_field}) one can split the time and space components of the stress-energy tensor and from that deduce the 3+1 matter terms entering Einstein's equations. 
Expressions of $E$, $P_i$ and $S_{ij}$, as a function of the scalar field components $R_\Phi$ and $I_\phi$ are given in Appendix \ref{s:matter}. 

Equations (\ref{e:hamilton_ok}-\ref{e:evol_ok}) are solved using the inner boundary conditions (\ref{e:full_set}) and demanding that, at spatial infinity, flat spacetime is recovered.

Concerning the Klein-Gordon equation, the situation is different from the linear case exposed in Sec. \ref{ss:hairy_clouds}. 
Indeed, with the coupling with gravity, the problem is no longer linear and admits solutions not for discrete values of the parameters but on a whole continuum. 
It follows that the whole procedure about $R_\Phi$ described in Sec. \ref{ss:hairy_clouds} is irrelevant. 
However, the discussion about the arbitrary phase in the definition of $R_\Phi$ and $I_\Phi$ still holds and the same procedure as in \ref{ss:hairy_clouds} is used.

The choice of the initial guess is of uttermost importance and that is where the knowledge of the cloud solutions of Sec. \ref{ss:hairy_clouds} is needed. 
Consider a linear solution of the Klein-Gordon equation, corresponding to a set of parameters. 
When moving away from this solution, by changing the value of some of the parameters, one can construct a sequence of non-linear solutions. 
The further away the parameters are from the linear values, the higher the amplitude of the scalar field. 
In this work, a sequence is constructed by varying the parameter $\Omega_{\rm BH}$, maintaining the equality $\omega = k \Omega_{\rm BH}$. 
This is only a choice and other parameters could be varied ($\mu$ for instance).

In order to use the cloud solution as an initial configuration, one proceeds as follows. 
A small amplitude for the maximum of the field is chosen and the cloud solution is scaled so that it has this maximum value. 
The parameter $\Omega_{\rm BH}$ is considered as an unknown of the numerical problem. 
By this it is to be understood that the solver is allowed to change its value. 
As the system contains an additional unknown it needs to be supplemented with an additional condition. 
This condition is the fixing of the amplitude of the field. 
By increasing the amplitude of the field, different configurations can be computed, corresponding to different values of $\Omega_{\rm BH}$. 
Alternatively, once the first full solutions are known, one can directly vary the parameter $\Omega_{\rm BH}$ along the sequence, in order to compute solutions with different amplitude of the field. 
The steps must be small enough so that the code does not converge to the trivial solution where the field vanishes everywhere.

\subsection{\label{ss:hairy_results} Numerical results}

A small modification in the numerical setting stems from the fact that the scalar field and black hole sizes are somewhat different. 
In the computations, the black hole is located at a coordinate radius of $r_{\rm H} =1$. 
However, the toroidal shape of the scalar field has a maximum at about 30 times this value. 
In order to deal with those two scales, more numerical domains than in Sec. \ref{s:kerr} are required. 
Typically, for each shell, the outer radius is twice the inner radius and a dozen of domains are used.

As already mentioned, a sequence of diverse values of $\Omega_{\rm BH}$ is exhibited. 
The starting point of the sequence is a cloud solution (see \ref{ss:hairy_clouds}) constructed from a Kerr black hole with $M \approx 5$ and $a/M \approx 0.95$, which corresponds to $\Omega_{\rm BH} \approx 0.0724$. 
The rotational quantum number is chosen to be $k=1$.

As in previous cases, one must check that the gauge quantities $K$ and $V^i$ converge to zero when resolution increases. 
The value of the expansion (\ref{e:expansion}) on the horizon should also go to zero. 
In order to do so, an arbitrary configuration is chosen and computed with different numbers of points. 
The chosen configuration corresponds to $\Omega_{\rm BH} \approx 0.069831$.

\begin{figure}
    \includegraphics[width=0.48\textwidth,keepaspectratio]{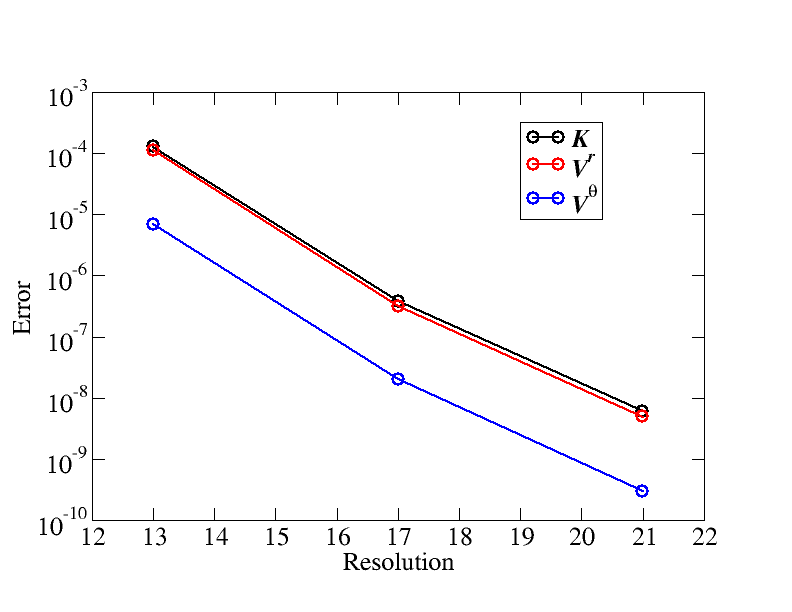}
    \includegraphics[width=0.48\textwidth,keepaspectratio]{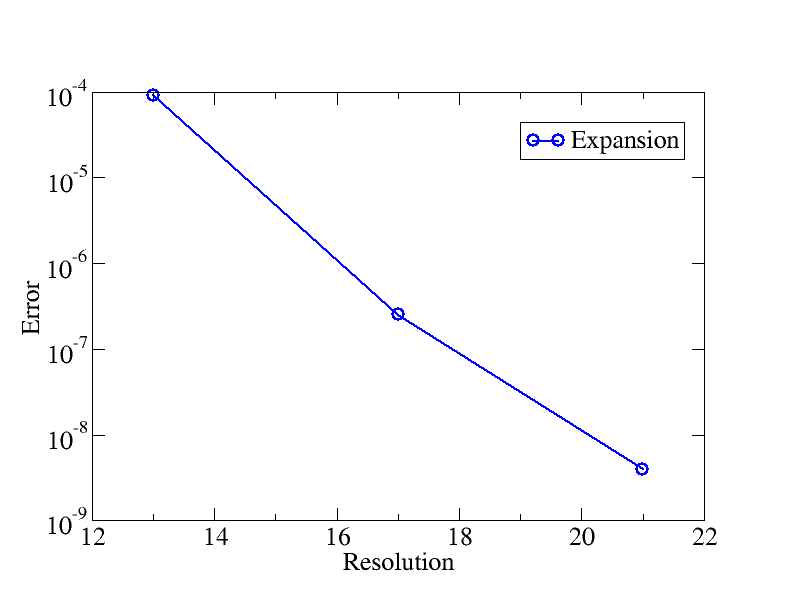}
    \caption{\label{f:errors_hairy} Maximal values of $K$ and $V^i$ (first panel) and of $\Theta$ on the horizon (second panel) as a function of resolution. The configuration emerges from a Kerr black hole with $M \approx 5$ and $a/M \approx 0.95$ and corresponds to $\Omega_{\rm BH} \approx 0.069831$. The various indicators exhibit a spectral convergence to zero.}
\end{figure}

The maximum values of $K$ and $V^i$ in the whole space and of $\Theta$ on the horizon are shown in Fig. \ref{f:errors_hairy}. 
The three different resolutions correspond to $\mathcal{N}_r = \mathcal{N}_\theta = 13$, $17$ and $21$, where $\mathcal{N}_r$ and $\mathcal{N}_\theta$ denote the number of collocation points in the $r$ and $\theta$ directions. 
The two panels of Fig. \ref{f:errors_hairy} clearly show a spectral convergence of the various quantities proving the validity of the solution.

As explained in Sec. \ref{ss:hairy_clouds}, the quantities $R_\Phi$ and $I_\Phi$ are defined up to a constant phase. 
The continuity of the radial derivative of $I_\Phi$ is relaxed, for one spectral coefficient. 
The matching of the radial derivative is replaced by a condition fixing the value of that same coefficient, for the field itself. 
It amounts to fixing the constant phase. 
In the following two possible choices are considered: one can demand that the coefficient of the field $I_\Phi$ vanishes (referred to as the zero phase condition) or that the coefficient of $I_\Phi$ is half that of $R_\Phi$ (referred to as the half phase condition). 

\begin{figure}
    \includegraphics[width=0.48\textwidth,keepaspectratio]{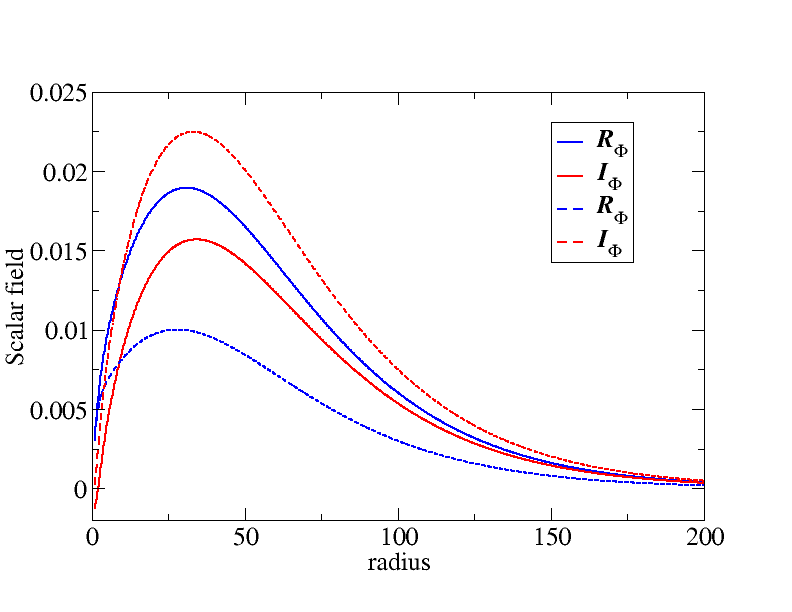}
    \includegraphics[width=0.48\textwidth,keepaspectratio]{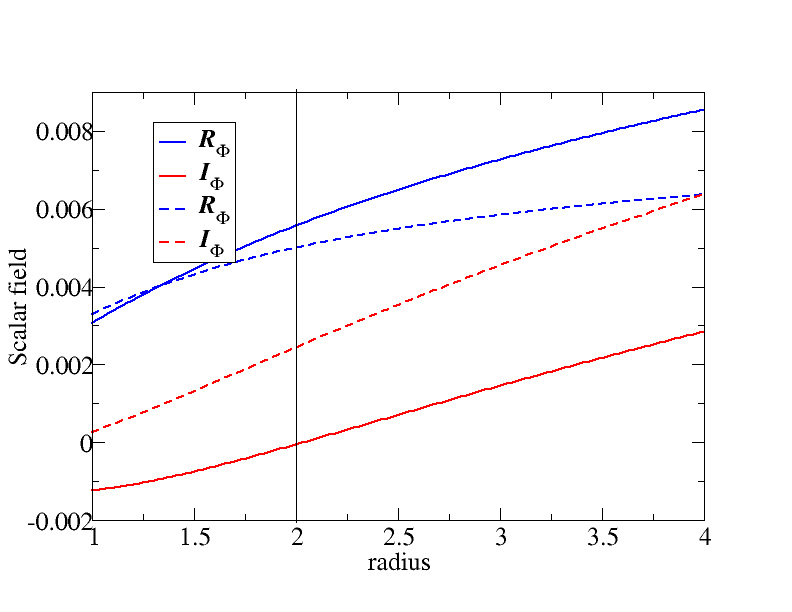}
    \caption{\label{f:phase_effect} Profiles along the $x$-axis of $R_\Phi$ (blue curves) and $I_\Phi$ (red curves) for the zero phase condition (solid lines) and the half phase one (dashed lines) (see text for details). The second panel shows the region close to the origin and the vertical line indicates the surface where the value of the phase is enforced.}
\end{figure}

For the solution with $\Omega_{\rm BH} = 0.067$, the fields resulting from those two cases are shown in Fig. \ref{f:phase_effect} where profiles of the fields, along the $x$-axis are shown. 
As expected the decomposition of the complex scalar field in terms of $R_\Phi$ and $I_\Phi$ changes. 
However, it is easy to check that the two relate by a constant phase rotation, the parameter being $\alpha \approx -0.464$ in that particular case. 
It is also easy to check that the two configurations lead to the same global quantities (mass, charge etc...) as they should.

\begin{figure}
\includegraphics[width=.7\textwidth,keepaspectratio]{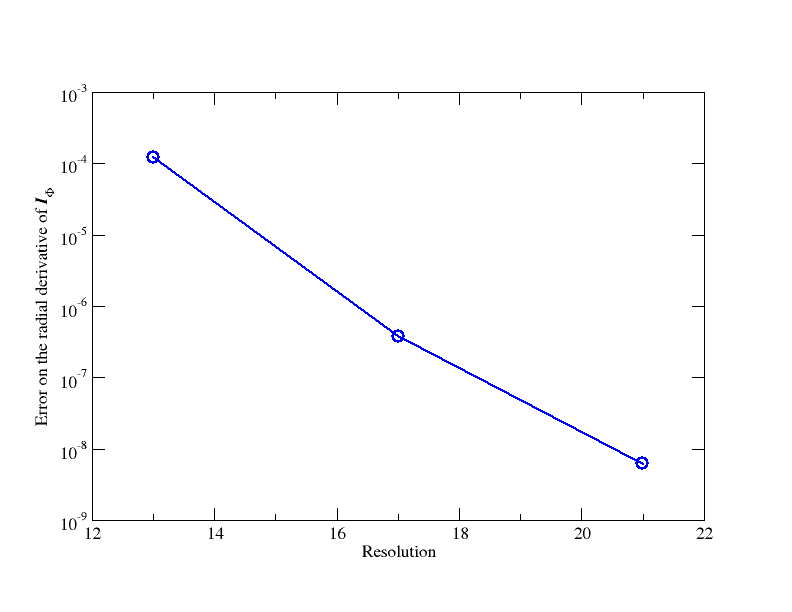}
\caption{\label{f:error_der_hairy} Discontinuity of the radial derivative of $I_\Phi$ as a function of resolution (i.e. the number of radial and angular coefficients). The configuration is the same as the one of Fig. \ref{f:errors_hairy}. }
\end{figure}

The continuity of the radial derivative of $I_\Phi$ can also be used to monitor the validity of the procedure. 
As it has not been enforced numerically, it needs to be checked that it is actually satisfied by the solution. 
Figure \ref{f:error_der_hairy} shows the error on the continuity of the radial derivative of $I_\Phi$, as a function of resolution. 
As expected with spectral methods, the error converges fastly to zero confirming that the solution is smooth as it should. 
Even if they are not explicitly shown here, the solutions passed several other tests (like equality of the ADM and Komar masses for instance).

\begin{figure}
    \includegraphics[width=0.48\textwidth,keepaspectratio]{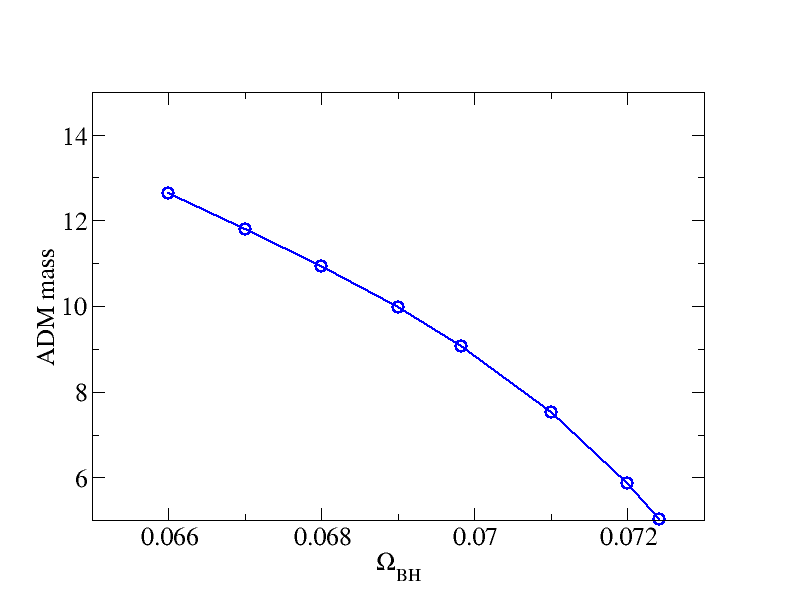}
     \includegraphics[width=0.48\textwidth,keepaspectratio]{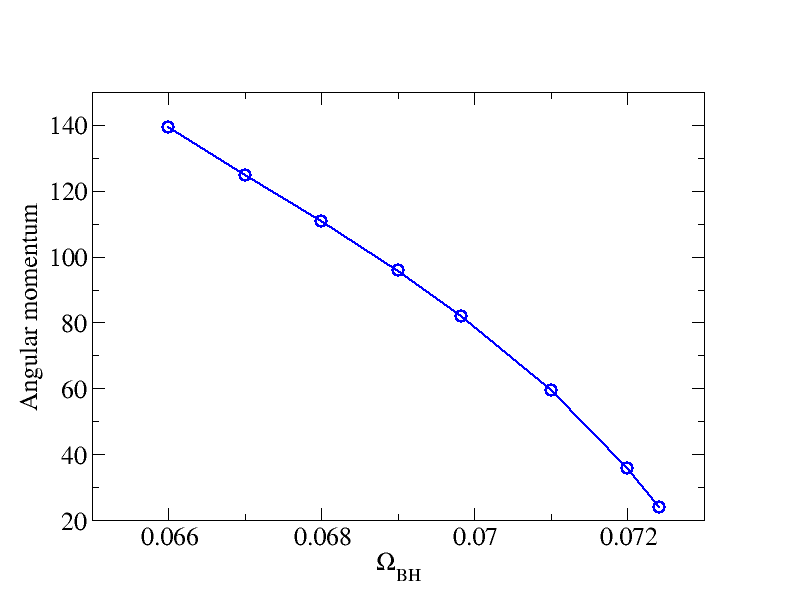}
    \caption{\label{f:global_hairy} ADM mass (first panel) and angular momentum (second panel), as a function of $\Omega_{\rm BH}$, for a sequence emerging from a Kerr black hole with $M \approx 5$ and $a/M \approx 0.95$. The Kerr black hole corresponds to the point at the right end of the curves.}
\end{figure}

The ADM mass and angular momentum of the solutions can be computed using the standard formulae (\ref{e:def_adm}) and (\ref{e:def_momentum}). 
As one moves along the sequence by going away from the Kerr black hole solution, those quantities increase, as the contribution of the scalar field to their values is more important. 
This can be seen in Fig. \ref{f:global_hairy}, where the mass and momentum are shown, as a function of $\Omega_{\rm BH}$. 
The last point to the right (i.e. with $\Omega_{\rm BH} \approx 0.0724$) corresponds to the Kerr black hole from which the sequence is constructed.

In order to monitor the relative importance of the black hole compared to the scalar field, following \cite{Herdeiro:2014, Herdeiro:2015}, one can compute the Noether charge. 
With the setting used in this paper it reads (see Sec. \ref{s:matter} for a precision about the use of the notations $\bar{R}_\Phi$ and $\bar{I}_\Phi$)

\bea
\label{e:noether}
Q &=& -\int_{\Sigma_t} \frac{1}{N} \l[\l(\omega + k \delta_i^\varphi B^i\r) \l(R_\Phi \bar{R}_\Phi + I_\Phi \bar{I}_\Phi\r)\r. \\
\nonumber
&+& \l. \frac{B^i}{2}\l(I_\Phi D_i \bar{R}_\Phi + \bar{I}_\Phi D_i R_\Phi - R_\Phi D_i \bar{I}_\Phi - \bar{R}_\Phi D_i I_\Phi\r)\r] \sqrt{\gamma} {\rm d}^3 x.
\eea
 
\begin{figure}
    \includegraphics[width=0.48\textwidth,keepaspectratio]{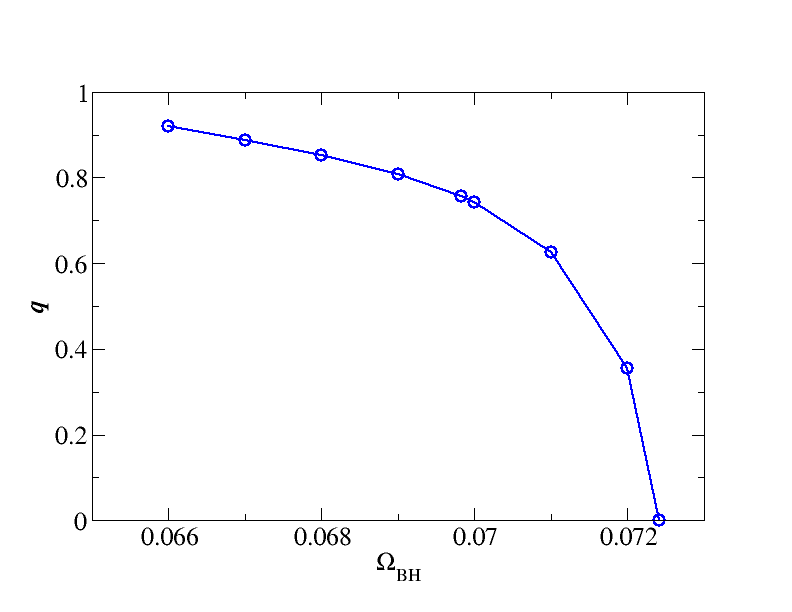}
     \includegraphics[width=0.48\textwidth,keepaspectratio]{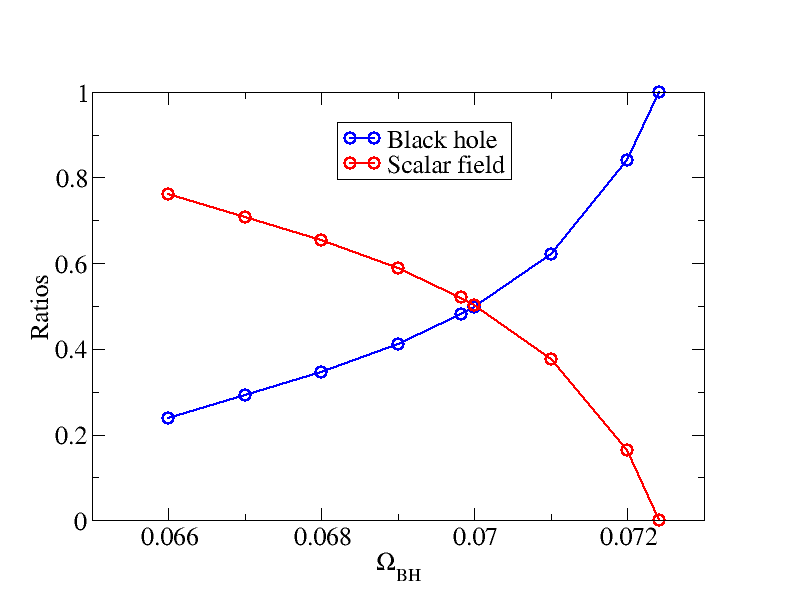}
    \caption{\label{f:ratios_field} The first panel shows the quantity $q$, as a function of $\Omega_{\rm BH}$. The second panel shows the ratios $M_{\rm BH} / M_{\rm ADM}$ and $M_\Phi / M_{\rm ADM}$, also as a function of $\Omega_{\rm BH}$. Both panels illustrate that one goes from a pure Kerr black hole situation (right side) to a situation where the field is dominant (left side).}
\end{figure}

 From the Noether charge one defines the quantity $q = k Q / J$. 
 For a Kerr black hole $q=0$ as the Noether charge vanishes. 
 For a boson star, it is known that $J = k Q$, so that $q=1$ \cite{Schunck:1996}.
 It follows that the quantity $q$ measures the relative importance of the hole and the field along the sequence, as it starts from $0$ for the Kerr black hole and approaches $1$ as the amplitude of the scalar field increases. 
 The quantity $q$ as a function of $\Omega_{\rm BH}$ is shown in the first panel of Fig. \ref{f:ratios_field}.
 
 The Komar mass can also be used as a way of measuring the influence of the field. 
 Indeed, if it can be computed as a surface integral at infinity by Eq. (\ref{e:def_komar}), it can also be obtained by the sum of two terms, $M_{\rm BH}$ a surface integral on the horizon and $M_\Phi$ a volume integral over matter terms. 
 The general expressions for $M_{\rm BH}$ and $M_\Phi$ can be found in \cite{Gourgoulhon:2007} (Eqs. (8.70) and (8.71)) respectively. 
 In the situation at hand they are obtained as

\bea
 \label{e:mbh}
M_{\rm BH} &=& \frac{1}{4\pi} \int_{r=r_{\rm h}} \sqrt{h} \l(\tilde{s}^i D_i N - K_{ij} \tilde{s}^i B^j\r) {\rm d} S \\
\label{e:mphi}
M_\Phi &=& \int_{\Sigma_t} \sqrt{\gamma} \l[N\l(E+\gamma^{ij} S_{ij}\r) - 2 P_i B^i\r] {\rm d}^3 x,
\eea
where $h$ denotes the determinant of the metric induced on the horizon $h_{ij} = \gamma_{ij} - \tilde{s}_i \tilde{s}_j$. 
In the second plot of Fig. \ref{f:ratios_field} those quantities are shown, as a function of $\Omega_{\rm BH}$. 
They are scaled by the total mass so that the sum is one (given that $M_{\rm Komar} = M_{\rm ADM}$). 
As for the quantity $q$, one goes from a regime dominated by the black hole to a situation where the scalar field contributes more to the total mass.

\begin{figure}
\includegraphics[width=.7\textwidth]{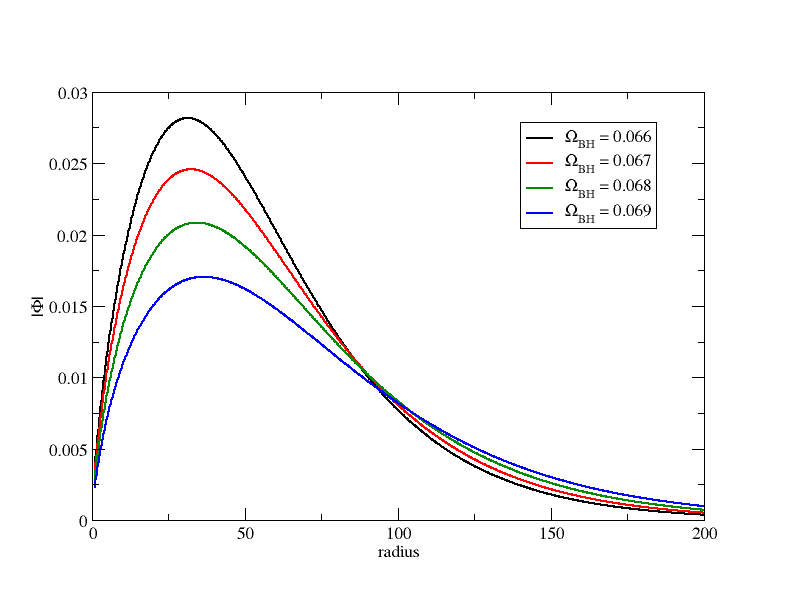}
\caption{\label{f:profiles_field} Values of $\l|\Phi\r|$ along the $x$-axis, for four different values of $\Omega_{\rm BH}$. The black hole horizon is located at $r_{\rm h}=1$.}
\end{figure}

Last, as an illustration, some profiles of $\l|\Phi\r| = \l(R_\Phi \bar{R}_\Phi + I_\Phi \bar{I}_\Phi\r)^{1/2}$ are shown in Fig. \ref{f:profiles_field}, along the $x$-axis. 
One can note that the maximum value of the field is located at about 40 times the radius of the black hole $r_{\rm H} = 1$. 
As expected, the further $\Omega_{\rm BH}$ is from the value of the Kerr black hole used to initiate the sequence, the higher the amplitude of the scalar field.

\section{\label{s:conclusions}Last words}

A formalism to compute spacetimes containing stationary black holes is presented. 
The geometry is described by the 3+1 decomposition of spacetime and the various metric fields are found using the maximal slicing gauge for the choice of time and the spatial harmonic gauge for the spatial coordinates. 
The presence of the hole itself is enforced by demanding that a given sphere is an apparent horizon. 
Moreover, as stationarity is assumed, one demands that this horizon is in equilibrium. 
Using a combination of analytical and numerical studies, a set of boundary conditions for the various metric quantities is found. 
Let us point out that this leads to a choice of coordinates that is regular everywhere, even on the horizon itself. 
This is to be contrasted with the analytic Boyer-Lindquist coordinates of the Kerr black hole, which are singular on the horizon. 
This property may prove useful in numerical applications.

The whole procedure is applied to three different situations. 
The system of equations is solved numerically using spectral methods and the \textsc{Kadath} library \cite{Grandclement:2009, Kadath}. 
The first and most simple application is to recover the classical Kerr black hole (in coordinates that are not analytical). 
Configurations up to a Kerr parameter $a/M$ of $0.99$ are easily computed. 
The second application concerns a static black hole where gravity is minimally coupled to a real scalar field, with a negative cosmological constant. 
It is called the MTZ black hole, from the names of the authors who obtained it analytically \cite{Martinez:2004}. 
The formalism appears to work well also in the case of this spacetime which is asymptotically anti-de Sitter. 
The last application is devoted to the computation of a family of black holes with complex scalar hairs \cite{Herdeiro:2014, Herdeiro:2015}. 
This last example combines a coupling with matter (i.e. the scalar field) and the inclusion of rotation. 
For the three cases, errors are carefully checked, especially by showing fast convergence to zero of several error indicators when the numerical resolution increases.

In the future, it is hoped that the formalism presented here will be a valuable tool to study various models of black holes. 
A lot of the results in the field of black holes in alternative theories of gravity relies, at some degree, on the analytic choice of coordinate systems. 
This can be a difficulty to get the more general solutions possible. 
This is especially true when rotation is included. 
For instance, there are no rotating equivalent of the MTZ black hole known yet.

Another extension of this work concerns black holes that are not in exact equilibrium.
In particular, this is the case for objects in binary systems. The boundary conditions 
proposed here could be applied to generate improved initial data for binary coalescence simulations. Usually initial data are 
generated assuming that the spatial metric is conformally flat (see \cite{Gourgoulhon:2001, Grandclement:2001, Caudill:2006, Papenfort:2021}).
This assumption could be relaxed using the techniques developed in this paper.
However, due to the emission of gravitational waves,
the horizons are not in equilibrium and thus it is not expected that the boundary conditions
will be exact. However their accuracy should improve as the separation increases. 
Equilibrium of the horizons are not the only place where deviation from stationarity must be accounted for.
The terms in $\partial_t$ present in the equations (\ref{e:hamilton}-\ref{e:evol}) must also be considered. 
Outer boundary conditions must also be carefully investigated, as simple spatial asymptotic flatness may not be sufficiently 
precise when gravitational waves are present. It is beyond the scope of this paper to implement the boundary conditions
presented here in the binary black hole context. However this is something that is planned for the future and one can hope
that the formalism exposed here will lead to significant improvement in the precision of the computation of initial data 
for binary black holes configurations.

\appendix

\section{\label{s:MTZ_system} System of equations for the MTZ black holes}
\subsection{\label{ss:MTZ_3+1} Matter terms for the MTZ black hole}

For a stationary real scalar field, the various 3+1 matter terms can be obtained as:
\be
    \label{e:3+1_rho}
    E = \dfrac{1}{2}D_i\phi D^i\phi + \dfrac{(B^i D_i\phi)^2}{2N^2} + V(\phi),
\ee
\be
    \label{e:3+1_p}
    P_i = \dfrac{(B^j D_j\phi)}{N}D_i\phi,
\ee
\be
    \label{e:3+1_S} 
    S_{ij} = T_{ij} = D_i\phi D_j\phi - \dfrac{1}{2}(D_k\phi D^k\phi) \gamma_{ij} + \dfrac{(B^k D_k\phi)^2}{2N^2} \gamma_{ij} - V(\phi)\gamma_{ij},
\ee
\be
    \label{e:3+1_TrS}
    S = -\dfrac{1}{2}(D_k\phi D^k\phi) + \dfrac{3}{2} \dfrac{(B^k D_k\phi)^2}{N^2} - 3V(\phi).
\ee

In this context, the Klein-Gordon equation is given by:
\be
    \label{e:3+1_KG}
    E_{KG} : \dfrac{1}{N}D_i\l(N\l(\gamma^{ij}-\dfrac{B^iB^j}{N^2}\r)D_j\phi\r) - \dfrac{dV}{d\phi} = 0.
\ee

\subsection{\label{ss:MTZ_reg} Regularization at the ADS border}

For the MTZ black hole, the regularized quantities used close to the ADS boundary are given. The various quantities are multiplied by the appropriate power of $\Omega$ (Eq. (\ref{e:MTZ_Omega})) in order to avoid divergences. It leads to

\bea
\Tilde{\gamma}^{ij} & = & \gamma^{ij}/\Omega^2 \\ 
\Tilde{B}_i & = & \Omega^2 B_i \\
\Tilde{D}_i\Tilde{\phi} & = & \Omega\partial_i\phi \\
\Tilde{D} & = & \Omega\partial \pm \Tilde{\Gamma}  \\
\Tilde{\Gamma}^k_{ij} = \Omega \Gamma^k_{ij} 
& = & \frac{1}{2} \Omega\Tilde{\gamma}^{kl}(\partial_i\Tilde{\gamma}_{lj} + \partial_j\Tilde{\gamma}_{il}- \partial_l\Tilde{\gamma}_{ij}) \\
\nonumber
 &-& \Tilde{\gamma}^{kl} (\Tilde{\gamma}_{lj}\partial_i\Omega + \Tilde{\gamma}_{il}\partial_j\Omega - \Tilde{\gamma}_{ij}\partial_l\Omega) \\ 
\Tilde{R}_{ij}  = \Omega^2 R_{ij} 
& = & \Omega(\partial\Tilde{\Gamma}^k_{ij} - \partial_i\Tilde{\Gamma}^k_{jk}) - (\Tilde{\Gamma}^k_{ij}\partial_k\Omega - \Tilde{\Gamma}^k_{jk}\partial_i\Omega) + \Tilde{\Gamma}^k_{ij}\Tilde{\Gamma}^l_{kl} - \Tilde{\Gamma}^l_{ik}\Tilde{\Gamma}^k_{jl} \\ 
\Tilde{R} & = & R \\
\widetilde{D_jN} = \Omega^2D_jN & = & \Tilde{D}_j\Tilde{N} - \Tilde{N}\partial_i\Omega \\ 
\widetilde{D_iD_jN} = \Omega^3 D_iD_jN & = & -2(\partial_i\Omega)\widetilde{D_jN} + \Tilde{D}_i\widetilde{D_jN} \\
\widetilde{D_iB_j} = \Omega^3D_iB_j & = & \Tilde{D}_i\Tilde{B}_j - 2\Tilde{B}_j\partial_i\Omega  \\
\Tilde{D}_i\Tilde{\gamma}_{jk} & = & 2\partial_i\Tilde{\gamma}_{jk} \neq 0 \\ \Tilde{D}_i\Tilde{\gamma}^{jk} & = & -2\partial_i\Omega\Tilde{\gamma}^{jk} \neq 0  \\
\Tilde{K}_{ij} = \Omega^2K_{ij} & = & (\widetilde{D_iB_j} + \widetilde{D_jB_i})/2\Tilde{N} \\
\Tilde{K}^i_j & = & K^i_j \\
\Tilde{K}^{ij} & = & K^{ij}/\Omega^2 \\ 
\Tilde{V}^k = V^k/\Omega & = & \Tilde{\gamma}^k_{ij} (\Tilde{\Gamma}^k_{ij} - \Tilde{\bar{\Gamma}}^k_{ij}) \\
\Tilde{V}_i & = & \Omega V_i  \\
\widetilde{D_iV_j} = \Omega^2D_iV_j & = & \Tilde{D}_i\Tilde{V}_j - \Tilde{V}_j\partial_i\Omega \\
\widetilde{\mathcal{L}_{\bm{B}}K}_{ij} = \Omega^3 \mathcal{L}_{\bm{B}}K_{ij} & = & \mathcal{L}_{\Tilde{\bm{B}}}\Tilde{K}_{ij} - 2 \Tilde{K}_{ij}\Tilde{B}^k\partial_k\Omega 
\eea

\bea
\Tilde{E} = E  & = & \dfrac{1}{2}\Tilde{\gamma}^{ij}\Tilde{D}_i\Tilde{\phi}\Tilde{D}_j\Tilde{\phi} + V(\Tilde{\phi}) + \dfrac{(\Tilde{B}^i\Tilde{D}_i\Tilde{\phi})^2}{2\Tilde{N}^2}  \\
\Tilde{P}_i = \Omega P_i & = & \dfrac{\Tilde{B}^i\Tilde{D}_i\Tilde{\phi}}{\Tilde{N}}\Tilde{D}_i \Tilde{\phi}  \\
\Tilde{S}_{ij} = \Omega^2 S_{ij} & = &\Tilde{D}_i\Tilde{\phi}\Tilde{D}_j\Tilde{\phi} - \dfrac{1}{2}(\Tilde{\gamma}^{kl}\Tilde{D}_k\Tilde{\phi}\Tilde{D}_l\Tilde{\phi})\Tilde{\gamma}_{ij} + \dfrac{(\Tilde{B}^i\Tilde{D}_i\Tilde{\phi})^2}{2\Tilde{N}^2}\Tilde{\gamma}_{ij} - V(\Tilde{\phi})\Tilde{\gamma}_{ij}  \\
\Tilde{S} = S & = & \Tilde{\gamma}^{ij}\Tilde{S}_{ij}
\eea

The regularized equations are thus:
\bea
\Tilde{H} = H  & : & \tilde{\gamma}^{kl} \l(\Tilde{R}_{kl} - \widetilde{D_k V_l}\r)  - \Tilde{K}_{ij}\Tilde{K}^{ij} - 2\Lambda - 16\pi G \Tilde{E} = 0 \\
\Tilde{M}_i = \Omega M_i & : & \Tilde{D}_j\Tilde{K}^j_i - 8\pi G\Tilde{P}_i =0 \\
\Tilde{E}_{ij} = \Omega^3E_{ij} & : & \widetilde{\mathcal{L}_{\Tilde{\bm{B}}}K}_{ij} - \widetilde{D_iD_jN} \\
\nonumber
&+& \Tilde{N} \l[ \Tilde{R}_{ij} - \widetilde{D_{(i}V_{j)}} - 2\Tilde{K}_{ik}\Tilde{K}^k_j - \Lambda\Tilde{\gamma}_{ij} -8\pi G \l( \Tilde{S}_{ij} - \dfrac{\Tilde{S} - \Tilde{E}}{2} \Tilde{\gamma}_{ij} \r) \r] = 0 \\
\widetilde{E_{KG}} = E_{KG} & : & \dfrac{1}{\Tilde{N}}\Tilde{D}_j\l( \Tilde{N} \l( \Tilde{\gamma}^{ij} - \dfrac{\Tilde{B}^i\Tilde{B}^j}{\Tilde{N}^2} \r) \Tilde{D}_i\Tilde{\phi} \r) = 0
\eea

\section{\label{s:matter}Matter terms for the Einstein-Klein-Gordon system}

The 3+1 matter terms derived from the ansatz (\ref{e:scalar_field}) are computed. Given the dependence in terms of $t$ and $\varphi$, one can compute the various following terms

\bea
\label{e:partialt}
\partial_t \Phi \partial_t \bar{\Phi} &=& \omega^2 \l(R_\Phi \bar{R}_\Phi + I_\Phi \bar{I}_\Phi\r)\\
\label{e:partialti}
\partial_t \Phi D_i \bar{\Phi} + \partial_t \bar{\Phi} D_i \Phi &=&
\omega\l(R_\Phi D_i \bar{I}_\Phi + \bar{R}_\Phi D_i I_\Phi - I_\Phi D_i \bar{R}_\Phi - \bar{I}_\Phi D_i R_\Phi\r) \\
\nonumber
&-& 2 k \omega \delta_i^\varphi \l(R_\Phi \bar{R}_\varphi + I_\Phi \bar{I}_\Phi\r)\\
\label{e:partialii}
D_i \Phi D_j \bar{\Phi} &=& D_i R_\Phi D_j \bar{R}_\Phi + D_i I_\Phi D_j \bar{I}_\Phi + k^2 \delta_i^\varphi \delta_j^\varphi \l(R_\Phi \bar{R}_\Phi + I_\Phi \bar{I}_\Phi\r) \\
\nonumber
&+& k \l(\delta_j^\varphi \bar{I}_\Phi D_i R_\Phi + \delta_i^\varphi I_\Phi D_j \bar{R}_\Phi - \delta_j^\varphi \bar{R}_\Phi D_i I_\Phi
- \delta_i^\varphi R_\Phi D_j \bar{I}_\Phi\r).
\eea

Even if $R_\Phi$ and $I_\Phi$ are real quantities, the notations $\bar{R}_\Phi$ and $\bar{I}_\Phi$ are kept to differentiate terms that are in  factor of $\exp\l[i\l(\omega t- k \varphi\r)\r]$ and those of  $\exp\l[-i\l(\omega t- k \varphi\r)\r]$.

The terms (\ref{e:partialt}-\ref{e:partialii}) enter into the expressions of the components of the stress-energy tensor as

\bea
T &\equiv& \nabla_\mu \Phi \nabla^\mu \bar{\Phi} +V\l(\l|\Phi\r|^2\r) =\\
\nonumber
&&-\frac{1}{N^2} \l[\partial_t \Phi \partial_t \bar{\Phi}\r] + \frac{B^i}{N^2} \l[\partial_t \Phi D_i \bar{\Phi} + \partial_t \bar{\Phi} D_i \Phi\r] + \l(\gamma^{ij} -\frac{B^i B^j}{N^2}\r) \l[D_i \Phi D_j \bar{\Phi}\r] + \mu^2 \l(R_\Phi \bar{R}_\Phi + I_\Phi \bar{I}_\Phi\r) \\
T_{tt} &=&  \l[\partial_t \Phi \partial_t \bar{\Phi}\r] - \frac{1}{2} \l(-N^2 +B_iB^i\r) T \\
T_{ti} &=& \l[\partial_t \Phi D_i \bar{\Phi} + \partial_t \bar{\Phi} D_i \Phi\r] - \frac{1}{2} B_i T \\
T_{ij} &=& \frac{1}{2} \l[D_i \Phi D_j \bar{\Phi}  + D_j \Phi D_i \bar{\Phi}\r] -\frac{1}{2} \gamma_{ij} T.
\eea

From those components, the 3+1 matter terms are expressed as
\bea
E &=& \frac{1}{N^2} T_{tt} - 2\frac{B^i}{N^2} T_{ti} + \frac{B^i B^j}{N^2} T_{ij} \\
P_i &=& -\frac{1}{N} T_{ti} + \frac{B^j}{N} T_{ij} \\
S_{ij} &=& T_{ij}.
\eea

\begin{acknowledgments}
The authors would like to thank Eric Gourgoulhon for many fruitful scientific discussions
on the topics covered by this paper.

This work was granted access to the HPC resources of MesoPSL financed
by the Region Ile de France and the project Equip@Meso (reference
ANR-10-EQPX-29-01) of the programme Investissements d’Avenir supervised
by the Agence Nationale pour la Recherche.
\end{acknowledgments}

\bibliography{mybib.bib}

\end{document}